\documentclass[a4paper,11pt]{article}
\usepackage{amssymb}
\usepackage{amsmath}
\usepackage{latexsym,amssymb,amsmath}
\usepackage{graphicx}
\usepackage{epsfig}

\DeclareFontFamily{U}{rsf}{} \DeclareFontShape{U}{rsf}{m}{n}{
  <5> <6> rsfs5 <7> <8> <9> rsfs7 <10-> rsfs10}{}
\DeclareMathAlphabet\Scr{U}{rsf}{m}{n} \makeatletter
\@addtoreset{equation}{section} \makeatother

\newcommand{\be}{\begin{equation}}
\newcommand{\ee}{\end{equation}}
\newcommand{\bea}{\begin{eqnarray}}
\newcommand{\eea}{\end{eqnarray}}
\newcommand{\ba}{\begin{array}}
\newcommand{\ea}{\end{array}}
\newcommand{\bit}{\begin{itemize}}
\newcommand{\eit}{\end{itemize}}
\newcommand{\ben}{\begin{enumerate}}
\newcommand{\een}{\end{enumerate}}

\begin{document}

\begin{titlepage}
 \thispagestyle{empty}
\begin{flushright}
     \hfill{CERN-PH-TH/2011-078}\\
 \end{flushright}

 \vspace{50pt}

 \begin{center}
     { \Huge{\bf      {Small Black Hole Constituents  \\
     \vspace{8pt}and Horizontal Symmetry
    }}}

     \vspace{25pt}

     {\Large {Anna Ceresole$^{a}$, Sergio Ferrara$^{b,c,d}$,\\ Alessio Marrani$^{b,c}$ and Armen Yeranyan$^{c,e}$}}

     \vspace{20pt}

     {\it ${}^a$ INFN, Sezione di Torino,\\
     Via Pietro Giuria 1, I-10125 Torino, Italy;\\
     \texttt{ceresole@to.infn.it}}

     \vspace{10pt}

  {\it ${}^b$ Physics Department, Theory Unit, CERN,\\
     CH -1211, Geneva 23, Switzerland;\\
     \texttt{sergio.ferrara@cern.ch}\\
     \texttt{alessio.marrani@cern.ch}}

     \vspace{10pt}

    {\it ${}^c$ INFN - Laboratori Nazionali di Frascati,\\
     Via Enrico Fermi 40, I-00044 Frascati, Italy\\
     \texttt{ayeran@lnf.infn.it}}

     \vspace{10pt}

     {\it ${}^d$  Department of Physics and Astronomy,\\
University of California, Los Angeles, CA 90095-1547,USA}\\

     \vspace{10pt}

{\it ${}^e$ Department of Physics, Yerevan State University\\
Alex Manoogian St. 1, Yerevan, 0025, Armenia}\\

     \vspace{10pt}

     \vspace{30pt}

     {ABSTRACT}

 \vspace{10pt}
 \end{center}
By exploiting the role of the horizontal symmetry $SL_{h}\left( 2,\mathbb{R}%
\right) $, we extend the analysis and classification of two-centered
extremal black hole charge configurations to the case of ``small''
single-centered constituents. These latter are seen to decrease the
number of independent horizontal-invariant polynomials from four to
one, depending on the rank of the charge orbit supporting each of
the two centers.
Within $U$-duality groups of type $E_{7}$, both reducible and
irreducible symmetric supergravity models in four space-time
dimensions are considered, thus encompassing $\mathcal{N}=2$ and
$\mathcal{N}=8$ theories.
\end{titlepage}
\tableofcontents
\section{\label{Intro}Introduction}

\textit{Multi-centered} black holes (BH) are a fascinating subject, dealing
with several aspects of quantum gravity, when this theory is defined through
supergravity, and its high-energy, fundamental completions, namely
superstrings and $M$-theory.
The discovery of the \textit{split attractor flow} and of \textit{walls of
marginal stability } \cite{D-1,D-2,BD} and the
corresponding issue of microstate BH counting \cite{DM-1}-\nocite{G}\cite{DM-2}, have been
some remarkable achievements in this field, also characterized by some
puzzling and yet not fully understood features, such as \textit{%
anti-marginal stability} and \textit{entropy enigmas} (see also \cite{G-1}-\nocite
{WC-1,Gimon-1,CS-1,David-1,Gaiotto-1,GP-1,WC-2,Pioline,Bena:2009ev}\cite{Gaiotto-2}; for
studies on $\mathcal{N}>2$, see \cite{FGK-1}-\nocite{Sen}\cite
{NB}). Earlier studies on composite (super)gravity solutions and
marginal stability, were done in \cite{Marginal-Refs}, while \cite
{Pioline-rev} provides a recent review on wall-crossing formul\ae .

This paper continues the investigation of geometric aspects of BH
physics, by exploiting  the (classical) duality symmetries of the
underlying supergravities \cite {CJ-1}, which are the continuum
limit of the U-duality \cite{HT-1} governing the non-perturbative
string dynamics, in the context of two-centered BH solutions. The
ultimate aim is to show that different aspects of multi-centered BH
dynamics are encoded into different values of (and constraints
among) certain multi-centered duality polynomial invariants. These
duality invariants characterize some multi-centered charge orbits,
which generalize the electric-magnetic charge orbits encoding  all
the main features of single-centered BH solutions, such as the
Bekenstein-Hawking entropy \cite{BH}, the ADM mass \cite{ADM} and
the BPS (supersymmetry-preserving) properties
\cite{FG1,LPS-1,CFM2,LA10-Proc}.

The single-centered orbits \cite{FG1} are known to provide a stratification of
the space of the irrep. $\mathbf{R}$ of the $d=4$  classical
$U$-duality group $G_{4}$ of the $\mathbf{f}$-dimensional
electric-magnetic charge vector $\mathcal{Q}\equiv \left( p^{\Lambda
},q_{\Lambda }\right) $ ($\Lambda =1,...,\mathbf{f}/2$).
Analogously, the $p$-centered orbits are expected to stratify the
space $\mathbf{R}_{1}\times\cdots\times \mathbf{R}_{p}$, for $p$  BH
constituents, with electro-magnetic fluxes  given by
\begin{equation}
\mathcal{Q}_{a}^M\equiv \left( p_a^{\Lambda },q_{a\, \Lambda }\right)\, \\ \ \ \ \ a=1,\ldots, p,\ \ \, ,\ \ M=1,\dots , \mathbf{f}.
\label{2-center}
\end{equation}

Interestingly, it has been recently uncovered \cite{FMOSY-1,Irred-1}
that  a $p\left( \geqslant 2\right) $-centered BH solution in $d=4$
space-time dimensions enjoys an ``horizontal'' symmetry
$SL_{h}\left( p,\mathbb{R}\right) $ among the centers of the BH
constituents. As a consequence of this symmetry, further invariant
polynomials in the charges  beside the usual $U$-duality invariants
acquire an important role, and they provide a tool to achieve a
finer classification of the allowed two-centered
configurations\footnote{ The horizontal symmetry has recently been
investigated within the fascinating connections with Quantum
Information Theory in \cite{Levay-2-center}}.

As commented in \cite{Irred-1}, the analysis can be performed for a generic number $p$ of centers, but, due to  the structure of the split flow in multi-center (super)gravity solutions \cite{D-1,D-2,BD,DM-1}, the case $p=2$  is already fully illustrative, at least regarding marginal stability. It is then natural to explore the   charge orbits for the horizontal doublet $%
\left( \mathcal{Q}_{1},\mathcal{Q}_{2}\right) $, and to
classify the allowed quantum numbers for the extremal BH two-centered compounds by  suitable invariant constraints among  horizontal invariant-polynomials.

For  $p=2$,
the simplest example of such ``horizontal invariants'' is provided by the
symplectic product of two charge vectors \cite{D-1,Irred-1}
\begin{equation}
\mathcal{W}\equiv \left\langle \mathcal{Q}_{1},\mathcal{Q}_{2}\right\rangle =%
\frac{1}{2}\mathcal{Q}_{a}^{M}\mathcal{Q}_{b}^{N}\mathbb{C}_{MN}\epsilon
^{ab},  \label{W}
\end{equation}
where  $a,b=1,2$, $\mathbb{C}_{MN}$ is the symplectic metric
\begin{equation}
\mathbb{C}_{MN}\equiv \left(
\begin{array}{cc}
0 & -\mathbb{I} \\
\mathbb{I} & 0
\end{array}
\right) ,  \label{sympl-metric}
\end{equation}
and $\epsilon $ is the usual $SL\left( 2,\mathbb{R}\right) \sim Sp\left( 2,%
\mathbb{R}\right) $ rank-$2$ invariant tensor.
By considering $\left( \mathcal{Q}%
_{1},\mathcal{Q}_{2}\right) $ as a doublet $\mathcal{Q}%
_{a}$ (spin $s=1/2$, fundamental irrep. $\mathbf{2}$) of $SL_{h}\left( 2,%
\mathbb{R}\right) $ \cite{FMOSY-1}, it is evident from (\ref{W}) that $%
\mathcal{W}$ is manifestly invariant under both the $U$-duality and
the horizontal symmetry. It is known that $\mathcal{W}$ enters the
description of many physical properties of the two-centered BH
compound states, such as the equilibrium distance between the two
centers, the intrinsic overall (orbital) angular momentum, and the
marginal stability condition \cite {D-1,BD}. Indeed, a crucial
feature of two-centered BH physics is that their physical properties
turn out to depend not only on $\mathcal{Q}_{1}+\mathcal{Q}_{2}$, as
for the BPS ADM mass \cite{MS-FM-1}, but also on other combination
of charges, such as the symplectic product (\ref{W}). Another
important instance is  the entropy at the (split) horizon, which is
the sum of the entropies of the two single-centered BH constituents
\cite{BD}:
\begin{equation}
\frac{S_{1+2}}{\pi }=\sqrt{\left| \mathcal{I}_{4}\left( \mathcal{Q}%
_{1}\right) \right| }+\sqrt{\left| \mathcal{I}_{4}\left( \mathcal{Q}%
_{2}\right) \right| },  \label{S-1+2}
\end{equation}
where $\mathcal{I}_{4}$ is the unique quartic invariant polynomial of the
irrep. $\mathbf{R}$ of the $U$-duality group $G_{4}$,
\begin{equation}
\mathcal{I}_{4}\left( \mathcal{Q}\right) =\frac{1}{2}\mathbb{K}_{MNPQ}\mathcal{Q}^{M}\mathcal{Q}^{N}\mathcal{Q}^{P}%
\mathcal{Q}^{Q},  \label{I4}
\end{equation}
and $\mathbb{K}_{MNPQ}$ is the so-called $K$-tensor \cite{Exc-Reds}.
As a natural two-centered generalization of the quartic invariant
$I_4 (\mathcal{Q})$, for $p=2$ one considers the contraction of the
$K$-tensor with four \textit{a priori} different charge vectors,
which gives rise to the symmetric  $\mathbf{I}_{abcd}$ tensor \cite
{FMOSY-1,Irred-1}, sitting in the spin $s=2$ irrep. $\mathbf{5}$ of $%
SL_{h}\left( 2,\mathbb{R}\right) $:
\begin{equation}
\mathbf{I}_{abcd} \equiv \frac{1}{2}\mathbb{K}_{MNPQ}\mathcal{Q}_{a}^{M}%
\mathcal{Q}_{b}^{N}\mathcal{Q}_{c}^{P}\mathcal{Q}_{d}^{Q}\, .
\end{equation}
If $G_4$ is a simple group, one can define the symmetric object in
the horizontal indices \cite{Irred-1}
\begin{equation}
T^\alpha_{(ab)}=t^\alpha_{MN} \mathcal{Q}^M_a \mathcal{Q}^N_b
\end{equation}
where $t_{~MN}^{\alpha }$ (${\alpha }=1,...,\mathbf{d}\equiv $dim$%
_{\mathbb{R}}\left( \mathbf{Adj}\left( G_{4}\right) \right) $) is
the symplectic representation of the generators of the Lie algebra
$\frak{g}_{4}$ of $G_4$. Using these tensors, one has
\begin{equation}
\mathbf{I}_{\left(abcd\right) } =-\frac{1}{6\tau }t_{~(MN}^{{\alpha }}t_{{\alpha }\mid PQ)}
\mathcal{Q}_{a}^{M}\mathcal{Q}_{b}^{N}\mathcal{Q}_{c}^{P}\mathcal{Q}
_{d}^{Q}=-\frac{1}{6\tau }T_{~(ab}^{{\alpha }}T_{{\alpha }
\mid cd)}
\end{equation}
 where $\tau $  is a model-dependent parameter \cite{Exc-Reds}:
\begin{equation}
\tau \equiv \frac{2\mathbf{d}}{\mathbf{f}\left( \mathbf{f}+1\right) }.\label%
{tau}
\end{equation}
A similar structure arises also when $G_4$ is a semisimple group
factorized as $SL(2,\mathbb R)\times SO(m,n)$, characterizing the
\textit{reducible} symmetric models listed in Table 1. Indeed,
within the so-called Calabi-Vesentini basis \cite{CV,CV-sugra}
symplectic frame, one can define the $\mathbb{T}$-tensor
\cite{FMOSY-1}
\begin{equation}
T_{(ab)[\Lambda\Sigma]}=p_{\Lambda (a} q_{ b)\Sigma}-q_{\Lambda
(a}p_{b)\Sigma}=\left(
\begin{array}{cc}
\mathbb{T}_{11} & \mathbb{T}_{12} \\
\mathbb{T}_{12} & \mathbb{T}_{22}
\end{array}
\right) _{\Lambda \Sigma },
\end{equation}
where indices are raised with the  the pseudo-Euclidean $SO(m,n)$
metric $\eta_{\Lambda\Sigma}$. This is a symmetric object in the
horizontal indices, whose components are the triplet of
$\mathbb{T}$-tensors of the \textit{reducible} models (called
${\mathbb T}_1, {\mathbb T}_2, {\mathbb T}_{12}$ in \cite{FMOSY-1}).
They constitute a two-centered generalization  of the product
$T_{\Lambda\Sigma}=p_\Lambda q_\Sigma -q_\Lambda p_\Sigma$, which
appears in  the single-centered fourth order invariant written as
\cite{CY,DLR,CT}
\begin{equation}
\mathcal{I}_{4}=\frac12
T_{\Lambda\Sigma}T_{\Gamma\Delta}\eta^{\Lambda\Gamma}\eta^{\Sigma\Delta}=-\frac{1}{2}\text{Tr}_{\eta
}\left( \mathbb{T}^{2}\right) =p^{2}q^{2}-\left( p\cdot q\right)
^{2} ,
\end{equation}
where ``Tr$_{\eta }$'' denotes the $\eta $-trace, namely the trace
in which the indices are contracted with $\eta $. For two
centers, in reducible models it generalizes to
\begin{equation}
\mathbf{I}_{abcd}=\frac{1}{2}T_{(ab\mid \Lambda \Sigma }T_{\mid
cd)\Gamma \Delta }\eta ^{\Lambda
\Gamma }\eta ^{\Sigma \Delta }=-\frac{1}{2}\text{Tr}_{\eta }\left( \mathbb{T}%
_{(ab}\mathbb{T}_{cd)}\right).
\end{equation}

The study of two-center extremal BH charge orbits associated to a generic horizontal doublet $\mathcal{Q}_1,\mathcal{Q}_2$ has been initiated in\cite{FMOSY-1,Irred-1}. It was found that in $d=4$ supergravity theories with symmetric scalar manifolds and for generic
%
charge vectors for each of the two centers, the dimension of a complete basis of $U$%
-duality invariant-polynomials is \textit{seven}, and it includes
both the horizontal singlet $\mathcal W$ and the quintet
$\mathbf{I}_{abcd}$. The minimum number of invariant polynomials
has been shown to decrease to \textit{four} if only polynomials
invariant under \textit{both} $U$-duality and horizontal symmetry
are taken into account.
The dimension of this $G_4\times SL_h(2,\mathbb{R})$ invariant basis further reduces in some specific cases, for instance, for some rank-$2$ and rank-$1$ symmetric scalar manifolds pertaining to the so-called $st^{2}$ and $t^{3}$, $%
\mathcal{N}=2$, $d=4$ models,  and for theories whose BH charge irrep. admits a
\textit{quadratic} invariant polynomial  $\left| \mathcal{I}_{2}\right| =\sqrt{%
\left| \mathcal{I}_{4}\right| }$, namely, $\mathcal{N}=2$ \textit{%
minimally coupled}\cite{Luciani} and  for $\mathcal{N}=3$ supergravity theories
\cite{MS-FMO-1}, which will not be dealt with in this paper.
 Interestingly,  the ``pure'' $\mathcal{N}=4$, $d=4$ supergravity, despite having a scalar manifold $\frac{SL\left( 2,%
\mathbb{R}\right) }{U\left( 1\right) }$ of rank $1$, has a complete
basis formed by seven duality invariants \cite{FMOSY-1}.\bigskip

The present investigation completes the previous analysis in that it determines  the restrictions and constraints on such invariants when at least one of the two centers is occupied by a  ``small'' black hole, having
$\mathcal{I}_{4}\left( \mathcal{Q}%
\right) =0$. Small black holes, corresponding to horizonless
solutions in Einstein two-derivatives (super)gravity, have zero entropy and they don't show an attractor behavior \cite{FG1}. However, they are an interesting sector of the BH
spectrum that has been recently  examined  in \cite{ADFT-FO-1,CFM2},
in particular regarding the different supersymmetry features of the
allowed orbits.

Large black hole charge orbits, with
$\mathcal{I}_{4}\neq 0$ are described by a minimum of four charges,
and in this sense the corresponding orbits are ``rank four''.
 More precisely,  the \textit{rank} of the orbit (or of the charge vector $\mathcal{Q}$ spanning it) is here the minimal number of charges which compose an orbit representative. Mathematically speaking, this defines the \textit{rank} of $\mathcal{Q}$ as element of the associated \textit{
Freudenthal triple system} \cite{Ferrar,Krut}. However, large orbits
can  become \textit{``lightlike''} when they satisfy  the condition
$\mathcal{I}_{4}=0$, and  their rank reduces to three. If a further
differential constraint $\partial \mathcal{I}_{4}/\partial
\mathcal{Q} =0$ is imposed, the rank further reduces to two
(\textit{critical} orbits), and it becomes one for
\textit{doubly-critical} orbits, having a suitable projection of
$\partial^2 \mathcal{I}_{4}/(\partial \mathcal{Q})^2$ vanishing
\cite{Ferrara-Maldacena}. In Sec. \ref{U-inv-constraints} we shall
revisit and add new results to the
manifestly $U$-invariant constraints defining the \textit{rank}, ranging from $4$ to $1$, of the single-centered charge vector $%
\mathcal{Q}$, which gives rise to the stratification of the representation space of the $U$-duality group.
When combining two centers, these constraints will clearly reflect in a number of combinations for the orbits of the compound system, which will be thoroughly investigated.


In Sec. \ref{2-Center-Relations} we shall deal with relations and properties of
the invariant polynomials characterizing a two-centered (extremal) BH
compound, which admit a natural interpretation and classification in terms
of the horizontal symmetry group $SL_{h}\left( 2,\mathbb{R}\right) $.

Then, the results of Secs. \ref{U-inv-constraints} and \ref
{2-Center-Relations} are used to perform a detailed analysis of all possible
two-centered (extremal) BH charge configurations, by considering all
possible combinations of the \textit{ranks} of the charge vectors $\mathcal{Q%
}_{1}$ and $\mathcal{Q}_{2}$ pertaining to the two single-centered BH
constituents.

\begin{table}[p]
\begin{center}
\begin{tabular}{|c||c|c|c|}
\hline $
\begin{array}{c}
\\
\mathcal{N} \\
~
\end{array}
$ & $\frac{G_{4}}{mcs\left( G_{4}\right) }$ & $\mathit{rank}$ & $
\begin{array}{c}
\\
J_{3} \\
\mathit{reducible}
\end{array}
$ \\ \hline\hline $
\begin{array}{c}
\\
2 \\
~
\end{array}
$ & $\frac{SL_{v}\left( 2,\mathbb{R}\right) }{U\left( 1\right) }\times \frac{%
SO\left( 2,n\right) }{SO\left( 2\right) \times SO\left( n\right)
},~n\in
\mathbb{N}~$ & $1+\text{min}\left( 2,n\right) $ & $\mathbb{R}\oplus \mathbf{%
\Gamma }_{1,n-1}~$ \\ \hline $
\begin{array}{c}
\\
4 \\
~
\end{array}
$ & $\frac{SL_{v}\left( 2,\mathbb{R}\right) }{U\left( 1\right) }\times \frac{%
SO\left( 6,n\right) }{SO\left( 6\right) \times SO\left( n\right)
},~n\in
\mathbb{N}\cup \left\{ 0\right\} ~$ & $1+\text{min}\left( 6,n\right) ~$ & $%
\mathbb{R}\oplus \mathbf{\Gamma }_{5,n-1}$ \\ \hline
\end{tabular}
\end{center}
\caption{\textit{Reducible~}symmetric~$d=4$ supergravity models.
``$mcs$'' stands for \textit{maximal compact subgroup} (with
symmetric embedding). The \textit{rank} of the scalar manifold, as
well as the related \textit{reducible} Euclidean rank-$3$ Jordan
algebra $J_{3}$ are also given (for further elucidation, see
\textit{e.g.} \protect\cite{ICL-rev} and Refs. therein). The
subscript ``$v$'' stands for \textit{``vertical''},
and it has been introduced in order to distinguish the $S$-duality $%
SL_{v}\left( 2,\mathbb{R}\right) $ group from the horizontal symmetry group $%
SL_{h}\left( 2,\mathbb{R}\right) $ }
\end{table}

\begin{table}[p]
\begin{center}
\begin{tabular}{|c||c|c|c|}
\hline $
\begin{array}{c}
\\
\mathcal{N}
\end{array}
$ & $\frac{G_{4}}{mcs\left( G_{4}\right) }~$ & \textit{rank} & $
\begin{array}{c}
\\
J_{3} \\
\mathit{irreducible}
\end{array}
$ \\ \hline\hline $
\begin{array}{c}
\\
2 \\
\left( t^{3}\text{ model}\right)
\end{array}
$ & $\frac{SL_{v}\left( 2,\mathbb{R}\right) }{U\left( 1\right) }$ & $1$ & $%
\mathbb{R}~$ \\ \hline $
\begin{array}{c}
\\
2 \\
~
\end{array}
$ & $\frac{Sp\left( 6,\mathbb{R}\right) }{U\left( 3\right) }~$ & $3~$ & $%
J_{3}^{\mathbb{R}}$ \\ \hline $
\begin{array}{c}
\\
2 \\
~
\end{array}
$ & $\frac{SU\left( 3,3\right) }{S\left( U(3)\times U\left( 3\right)
\right) }$ & $3$ & $J_{3}^{\mathbb{C}}~$ \\ \hline $
\begin{array}{c}
\\
2\overset{\text{``twin''}}{\leftrightarrow }6 \\
~
\end{array}
$ & $\frac{SO^{\ast }\left( 12\right) }{U\left( 6\right) }~$ & $3~$ & $%
J_{3}^{\mathbb{H}}~$ \\ \hline $
\begin{array}{c}
\\
2 \\
~
\end{array}
$ & $\frac{E_{7\left( -25\right) }}{E_{6\left( -78\right) }\times
U\left( 1\right) }$ & $3$ & $J_{3}^{\mathbb{O}}~$ \\ \hline $
\begin{array}{c}
\\
5 \\
~
\end{array}
$ & $\frac{SU\left( 1,5\right) }{U\left( 5\right) }$ & $1$ &
$M_{1,2}\left( \mathbb{O}\right) $ \\ \hline $
\begin{array}{c}
\\
8 \\
~
\end{array}
$ & $\frac{E_{7\left( 7\right) }}{SU\left( 8\right) }$ & $7$ & $J_{3}^{%
\mathbb{O}_{s}}$ \\ \hline
\end{tabular}
\end{center}
\caption{\textit{Irreducible~}symmetric~$d=4$ supergravity models. $\mathcal{%
N}=2$ magical quaternionic Maxwell-Einstein supergravity and
$\mathcal{N}=6$ ``pure'' supergravity are \textit{``twin''}, namely
they share the same bosonic sector \protect\cite
{ADF-fixed,FGimK,ADFGT,Samtleben,Ferrara-Duff-last-1}.
$M_{1,2}\left( \mathbb{O}\right) $ is the Jordan triple system (not
upliftable to $d=5$) generated by $2\times 1$ matrices over
$\mathbb{O}$ \protect\cite{GST}. Note that, with the exception of
the \textit{reducible} - but \textit{triality symmetric} - $stu$
model \protect\cite{DLR,BKRSW}, \textit{irreducible} models are all
ones for which the treatment of \protect\cite{Exc-Reds} holds (see
\textit{e.g.} Table 1 therein)}
\end{table}

This analysis is carried out for all supergravity theories
(with symmetric scalar manifold) whose $d=4$ $U$-duality group $G_{4}$ is a
\textit{``group of type }$E_{7}$\textit{''}, namely a group with a
symplectic representation $\mathbf{R}$ admitting a completely symmetric rank-%
$4$ invariant structure $\mathbf{q}$ such that the invariant polynomial $%
\mathcal{I}_{4}$ can be defined as\footnote{%
The normalization of $\mathbf{q}$ used here is the same as in \cite
{Duff-Freud-1}, and thus it differs by a factor $2$ with respect to the one
adopted \textit{e.g.} in \cite{Brown-E7}, \cite{Exc-Reds} and \cite{Irred-1}%
. The same holds \textit{e.g.} for Eq (\ref{spin=3/2}) further below.} \cite{Brown-E7}
\begin{eqnarray}
\mathcal{I}_{4}\left( \mathcal{Q}\right) &\equiv &\frac{1}{2}\left. \mathbf{q%
}\left( \mathbf{q}_{1},\mathbf{q}_{2},\mathbf{q}_{3},\mathbf{q}%
_{4}\right) \right| _{\mathbf{q}_{1}=\mathbf{q}_{2}=\mathbf{q}_{3}=%
\mathbf{q}_{4}\equiv \mathcal{Q}}  \notag \\
&\equiv&\frac{1}{2}\mathbb{K}_{MNPQ}\mathcal{Q}^{M}\mathcal{Q}^{N}\mathcal{Q}^{P}%
\mathcal{Q}^{Q}\, .
\end{eqnarray}
It is also worth recalling that the ``group of type $E_{7}$'' $G_{4}$ is a
symmetry group of Jordan algebra-related structures, namely:
\begin{equation}
G_{4}\sim Aut\left( \frak{M}\left( J_{3}\right) \right) \sim Conf\left(
J_{3}\right) ,  \label{G_4}
\end{equation}
where $Aut\left( \frak{M}\left( J_{3}\right) \right) $ is the automorphism
group of the vector space $\frak{M}\left( J_{3}\right) \equiv \mathbb{R}%
\oplus \mathbb{R}\oplus J_{3}\oplus J_{3}$ constructed over the Euclidean
rank-$3$ Jordan algebra $J_{3}$, whose conformal group is $Conf\left(
J_{3}\right) $ (see \textit{e.g.} \cite{Gunaydin-Lectures,ICL-1} for recent
reviews and lists of Refs.).
On the other hand, the $G_{4}$'s of \textit{minimally coupled} $\mathcal{N}=2$ , and $\mathcal{N}=3$ \cite{N=3} supergravities, omitted in this investigation, do not enjoy an interpretation in terms of (rank-$3$ Euclidean) Jordan algebras.

The groups of type $E_{7}$ $G_{4}$'s (which are $U$-duality groups
of supergravity theories in $d=4$ space-time dimensions with
symmetric scalar manifolds) may be grouped into two classes,
depending on whether $G_{4}$ is a semisimple Lie group, or it is a
simple Lie group itself.
The former case, analyzed in Sec. \ref{Red-Analysis}, corresponds to the
\textit{reducible} symmetric models, whose scalar manifolds\footnote{%
In matter coupled theories, we consider vector multiplets' scalar manifolds;
for instance, this is the case for all models of Table 1, with the exception
of ``pure'' $\mathcal{N}=4$ supergravity.} are grouped into two infinite
sequences, reported in Table 1. The first ($n=1$) element of the $\mathcal{N}%
=2$ sequence in Table 1, namely the so-called $st^{2}$ model, is \textit{%
non-generic}, and it deserves a separate treatment, given in Sec.
\ref {Red-Analysis-st^2}. The corresponding two-centered (extremal) BH charge
orbits  with \textit{both}
$\mathcal{Q}_{1}$ and $\mathcal{Q}_{2}$ ``large'' have been studied
in \cite{FMOSY-1}. Generally, the number of independent
$G_{4}$-invariant is seven, and a complete basis can be taken to be
\cite{FMOSY-1}
\begin{equation}
\text{~}\mathcal{W},~\mathcal{X},~\mathbf{I}_{abcd},
\label{p=2-red-duality-basis}
\end{equation}
where the quartic polynomial $\mathcal{X}$ is defined by Eq. (\ref{Xcall-red}%
).

The latter case, analyzed in Sec. \ref{Irred-Analysis}, corresponds to the
\textit{irreducible} symmetric models, namely to the so-called $\mathcal{N}%
=2 $ $t^{3}$ model, to the $\mathcal{N}=2$ magical Maxwell-Einstein
supergravity theories as well as to $\mathcal{N}=5$, $6$, $8$ ``pure''
supergravities, whose scalar manifolds are reported in Table 2. The $t^{3}$
model is \textit{non-generic} and it deserves a separate treatment, given in
Sec. \ref{Irred-Analysis-t^3}.

The two-centered (extremal) BH ``large'' charge orbits
of \textit{generic} irreducible symmetric models and of $t^{3}$ model have been
studied respectively in \cite{Irred-1} and in \cite{FMOSY-1}.
Generically, the number of independent $G_{4}$-invariant is seven, and a
complete basis can be taken to be \cite{Irred-1}
\begin{equation}
\mathcal{W},~\mathbf{I}_{abcd},~\mathbf{I}_{6},
\label{p=2-irred-duality-basis}
\end{equation}
where the sextic polynomial $\mathbf{I}_{6}$ is defined by Eq.
(\ref{rel-1}). The main
difference between the sets (\ref{p=2-red-duality-basis}) and
(\ref{p=2-irred-duality-basis}) is that $\mathcal{X}=0$
(\ref{Xcall-irred}) in irreducible cases, where also the constraint
(\ref{P12=0}) of degree twelve in the charges does not hold. It will
be emphasised in Sec. \ref{2-Center-Relations} that the existence of
$\mathcal{X}$ in reducible models can be traced back to the
\textit{semi-simple} nature of the $U$-duality group $G_{4}$, giving
rise to two independent quartic polynomials $\mathbf{I}^{\prime }$
and $\mathbf{I}^{\prime \prime }$ with vanishing horizontal
helicity, and related to $\mathbf{I}_{0}$ by Eq. (\ref
{rel-I0-I'-I''}).

The outcome of this analysis is that the number of independent duality- and horizontal- invariant
polynomials is decreased when one or both charge vectors are \textit{%
``small''}, since the vanishing of the quartic form (\ref{I4})  for one
center or both gives rise to various polynomial relations among otherwise independent invariants.
For example, as yielded by the analysis of Secs. \ref
{Red-Analysis} and \ref{Irred-Analysis}, if at least one of the two charge
vectors $\mathcal{Q}_{1}$ and $\mathcal{Q}_{2}$ is \textit{doubly-critical} (%
\textit{i.e.} rank $1$, see Sec. \ref{U-inv-constraints}; in the $\mathcal{N}%
=8$ language, this is $\frac{1}{2}$-BPS), then only one
\textit{independent} horizontal invariant, say the symplectic
product $\mathcal{W}$ (\ref{W}), exists.  This is one of the few
examples known in the literature \cite{BD}.

Notice that the results concerning the $%
st^{2}$ and $t^{3}$ models are consistent with the \textit{``}$%
stu\rightarrow st^{2}\rightarrow t^{3}$\textit{\ reduction''} (see \textit{%
e.g.} the discussion in \cite{BFGM1,BMOS-1,FMOSY-1}, and the recent analysis
in \cite{Levay-2-center}).

Finally, in Sec. \ref{Comment} we make some comments on some ``small $+$
small'' two-centered charge configurations corresponding to $\mathcal{I}%
_{4}\left( \mathcal{Q}_{1}+\mathcal{Q}_{2}\right) <0$, \textit{i.e.} to a BH
compound that, \textit{regarded as a single-centered solution}, is a
``large'' non-BPS (extremal) BH, making contact with recent literature, such
as \cite{CS-1} and \cite{Sen-2-center}.


\section{\label{U-inv-constraints}$U$-Invariant Constraints on $\mathcal{Q}$}

We start by giving a \textit{r\'{e}sum\'{e}} (with original results) of the $%
U$-invariant constraints defining the charge orbits of a single-center
(extremal) BH, namely of the $G_{4}$-invariant conditions defining the
\textit{rank} of the dyonic charge vector $\mathcal{Q}\in \mathbf{R}$ as an
element of the corresponding Freudenthal triple system (FTS) (see \cite
{Ferrar,Krut}, and Refs. therein); as mentioned above, $G_{4}$ is the $d=4$ $%
U$-duality group ``of type $E_{7}$'' \cite{Brown-E7}, and $\mathbf{R}$ is
its relevant BH charge irrep.. The symplectic indices $M=1,...,\mathbf{f}$ ($%
\mathbf{f}\equiv $dim$_{\mathbb{R}}\mathbf{R}\left( G_{4}\right) $) are
raised and lowered with the symplectic metric $\mathbb{C}_{MN}$ defined by (%
\ref{sympl-metric}). By recalling the definition (\ref{I4}) of the unique
(quartic) $G_{4}$-invariant polynomial constructed with $\mathcal{Q}\in
\mathbf{R}$, the \textit{rank} of a non-null $\mathcal{Q}$ as an element of $%
FTS\left( G_{4}\right) $ range from four to one, and it is manifestly $G_{4}$%
-invariantly characterized as follows:

\begin{enumerate}
\item  \textit{rank}$\left( \mathcal{Q}\right) =4$:
``large'' extremal BHs, with non-vanishing area of the event horizon:
\begin{equation}
\mathcal{I}_{4}\left( \mathcal{Q}\right) \gtrless 0.  \label{rank=4}
\end{equation}

\item  \textit{rank}$\left( \mathcal{Q}\right) =3$:
``small'' \textit{lightlike} extremal BHs, with vanishing area of the event
horizon:
\begin{eqnarray}
\mathcal{I}_{4}\left( \mathcal{Q}\right) &=&0;  \label{rank=3-1} \\
\frac{\partial \mathcal{I}_{4}}{\partial \mathcal{Q}^{M}} &\neq
&0\Leftrightarrow \mathbb{K}_{MNPQ}\mathcal{Q}^{N}\mathcal{Q}^{P}\mathcal{Q}%
^{Q}\neq 0,\text{\textit{at least}~for some }M\text{.}  \label{rank=3-2}
\end{eqnarray}

\item  \textit{rank}$\left( \mathcal{Q}\right) =2$:
``small'' \textit{critical} extremal BHs:
\begin{eqnarray}
\frac{\partial \mathcal{I}_{4}}{\partial \mathcal{Q}^{M}} &=&0%
\Leftrightarrow \mathbb{K}_{MNPQ}\mathcal{Q}^{N}\mathcal{Q}^{P}\mathcal{Q}%
^{Q}=0,~\forall M;  \label{rank=2-1} \\
\left. \frac{\partial ^{2}\mathcal{I}_{4}}{\partial \mathcal{Q}^{M}\partial
\mathcal{Q}^{N}}\right| _{\mathbf{Adj}\left( G_{4}\right) } &\neq &0,\text{%
\textit{at least}~for some }M,N\text{.}  \label{rank=2-2}
\end{eqnarray}

\item  \textit{rank}$\left( \mathcal{Q}\right) =1$:
``small'' \textit{doubly-critical} extremal BHs \cite
{Ferrara-Maldacena,DFL-0-brane}:
\begin{equation}
\left. \frac{\partial ^{2}\mathcal{I}_{4}}{\partial \mathcal{Q}^{M}\partial
\mathcal{Q}^{N}}\right| _{\mathbf{Adj}\left( G_{4}\right) }=0\Leftrightarrow
\left( 3\mathbb{K}_{MNPQ}+\mathbb{C}_{MP}\mathbb{C}_{NQ}\right) \mathcal{Q}%
^{P}\mathcal{Q}^{Q}=0,\text{ }\forall M,N.  \label{rank=1}
\end{equation}
\end{enumerate}


Let us consider the doubly-criticality condition (\ref{rank=1}) more in
detail. \textit{At least} for ``groups of type $E_{7}$'' \cite{Brown-E7}
which are $U$-duality groups $G_{4}$ of \textit{irreducible} symmetric
models in $d=4$, it holds that
\begin{eqnarray}
\left( \mathbf{R}\times \mathbf{R}\right) _{s} &=&\mathbf{Adj}+\mathbf{S};
\label{symm} \\
\left( \mathbf{R}\times \mathbf{R}\right) _{a} &=&\mathbf{1}+\mathbf{A},
\label{skew-symm}
\end{eqnarray}
where the subscripts ``$s$'' and ``$a$'' clearly stand for symmetric and
skew-symmetric. The presence of the singlet (which is nothing but $\mathbb{C}%
_{MN}$ defined in (\ref{sympl-metric})) in the skew-symmetric part (\ref
{skew-symm}) characterizes the BH charge irrep. $\mathbf{R}$ to be \textit{%
symplectic}. For example, for $G_{4}=E_{7}$ ($\mathbf{R}=\mathbf{56}$, $%
\mathbf{Adj}=\mathbf{133}$) one gets (see \textit{e.g.}
\cite{Slansky}; the subscripts "$s$" and "$a$" respectively stand
for symmetric and antisymmetric)
\begin{eqnarray}
\left( \mathbf{56}\times \mathbf{56}\right) _{s} &=&\mathbf{133}+\mathbf{1463%
}; \\
\left( \mathbf{56}\times \mathbf{56}\right) _{a} &=&\mathbf{1}+\mathbf{1539}.
\end{eqnarray}
For such groups, one can construct the projector operator on $\mathbf{Adj}%
\left( G_{4}\right) $:
\begin{eqnarray}
\mathcal{P}_{AB}^{~~CD} &=&\mathcal{P}_{\left( AB\right) }^{~~\left(
CD\right) }; \\
\mathcal{P}_{AB}^{~~CD}\frac{\partial ^{2}\mathcal{I}_{4}}{\partial \mathcal{%
Q}^{C}\partial \mathcal{Q}^{D}} &=&\left. \frac{\partial ^{2}\mathcal{I}_{4}%
}{\partial \mathcal{Q}^{A}\partial \mathcal{Q}^{B}}\right| _{\mathbf{Adj}%
\left( G_{4}\right) }; \\
\mathcal{P}_{AB}^{~~CD}\mathcal{P}_{CD}^{~~EF}\frac{\partial ^{2}\mathcal{I}%
_{4}}{\partial \mathcal{Q}^{E}\partial \mathcal{Q}^{F}} &=&\mathcal{P}%
_{AB}^{~~EF}\frac{\partial ^{2}\mathcal{I}_{4}}{\partial \mathcal{Q}%
^{E}\partial \mathcal{Q}^{F}},
\end{eqnarray}
where (recall (\ref{symm}))
\begin{eqnarray}
\frac{\partial ^{2}\mathcal{I}_{4}}{\partial \mathcal{Q}^{A}\partial
\mathcal{Q}^{B}} &=&\left. \frac{\partial ^{2}\mathcal{I}_{4}}{\partial
\mathcal{Q}^{A}\partial \mathcal{Q}^{B}}\right| _{\mathbf{Adj}\left(
G_{4}\right) }+\left. \frac{\partial ^{2}\mathcal{I}_{4}}{\partial \mathcal{Q%
}^{A}\partial \mathcal{Q}^{B}}\right| _{\mathbf{S}\left( G_{4}\right) }; \\
\left. \frac{\partial ^{2}\mathcal{I}_{4}}{\partial \mathcal{Q}^{A}\partial
\mathcal{Q}^{B}}\right| _{\mathbf{Adj}\left( G_{4}\right) } &=&2\left(
1-\tau \right) \left( 3\mathbb{K}_{ABCD}+\mathbb{C}_{AC}\mathbb{C}%
_{BD}\right) \mathcal{Q}^{C}\mathcal{Q}^{D}; \\
\left. \frac{\partial ^{2}\mathcal{I}_{4}}{\partial \mathcal{Q}^{A}\partial
\mathcal{Q}^{B}}\right| _{\mathbf{S}\left( G_{4}\right) } &=&2\left[ 3\tau
\mathbb{K}_{ABCD}+\left( \tau -1\right) \mathbb{C}_{AC}\mathbb{C}_{BD}\right]
\mathcal{Q}^{C}\mathcal{Q}^{D},
\end{eqnarray}
where the model-dependent parameter $\tau $ is defined by (\ref{tau}). The
explicit expression of $\mathcal{P}_{AB}^{~~CD}$ reads\footnote{%
For related results in terms of a map formulated in the ``$4D/5D$ special
coordinates'' symplectic frame (and thus manifestly covariant under the $d=5$
$U$-duality group $G_{5}$), see \textit{e.g.} \cite{Shukuzawa,Yokota}.} ($%
\alpha =1,...,\mathbf{d}$):
\begin{equation}
\mathcal{P}_{AB}^{~~CD}=\tau \left( 3\mathbb{C}^{CE}\mathbb{C}^{DF}\mathbb{K}%
_{EFAB}+\delta _{(A}^{C}\delta _{B)}^{D}\right) =-t^{\alpha \mid
CD}t_{\alpha \mid AB},  \label{P-Adj}
\end{equation}
where the relation \cite{Exc-Reds} (see also \cite{Og-1})
\begin{equation}
\mathbb{K}_{MNPQ}=-\frac{1}{3\tau }t_{(MN}^{\alpha }t_{\alpha \mid PQ)}=-%
\frac{1}{3\tau }\left[ t_{MN}^{\alpha }t_{\alpha \mid PQ}-\tau \mathbb{C}%
_{M(P}\mathbb{C}_{Q)N}\right] ,  \label{rel-2}
\end{equation}
where
\begin{equation}
t_{MN}^{\alpha }=t_{\left( MN\right) }^{\alpha };~~t_{MN}^{\alpha }\mathbb{C}%
^{MN}=0
\end{equation}
is the symplectic representation of the generators of the Lie algebra $\frak{%
g}_{4}$ of $G_{4}$. Notice that $\tau <1$ defined in (\ref{tau}) is just the ratio of the dimensions of the adjoint $\mathbf{Adj}$ and rank-$2$
symmetric $\left( \mathbf{R}\times \mathbf{R}\right) _{s}$ (\ref{symm})
reps. of $G_{4}$, or equivalently the ratio of upper and lower indices of $%
t_{MN}^{\alpha }$'s themselves. It should also be noted that, with respect
to the treatment given in \cite{Exc-Reds}, the result (\ref{rel-2}) has been
supplemented with the relation $\xi =-\frac{1}{3\tau }$ \cite{Irred-1},
obtained as a \textit{consistency condition} within the computations
yielding to (\ref{P-Adj}).

The result (\ref{P-Adj}) is a direct consequence of the fact that $%
t_{MN}^{\alpha }$ is the projector of $\left( \mathbf{R}\times \mathbf{R}%
\right) _{s}$ onto $\mathbf{Adj}$ (recall (\ref{symm})). More precisely, it
holds that
\begin{equation}
\mathcal{P}_{AB}^{~~CD}t_{CD}^{\alpha }=t_{AB}^{\alpha },
\end{equation}
where the normalization (see \textit{e.g.} Eq. (2.5) of \cite{Exc-Reds})
\begin{equation}
t_{~F}^{\alpha ~C}t_{~C}^{\beta ~F}=g^{\alpha \beta }
\end{equation}
has been used.

\section{\label{2-Center-Relations}Two-Centered Relations}

In order to study multi-centered charge configurations, it is worth
considering some general relations for $p$($\geqslant 2$)-center
(extremal)
BHs, which are manifestly covariant under the horizontal symmetry $%
SL_{h}\left( p,\mathbb{R}\right) $ introduced in \cite{FMOSY-1}; we will
here focus on the case $p=2$.

From \cite{FMOSY-1,Irred-1} and the Introduction, we recall $\mathbf{I}%
_{abcd}$ and
\begin{equation}
\widetilde{\mathcal{Q}}_{M\mid abc}\equiv \frac{1}{4}\frac{\partial \mathbf{I%
}_{abcd}}{\partial \mathcal{Q}_{d}^{M}}=\frac{1}{2}\mathbb{K}_{MNPQ}\mathcal{%
Q}_{a}^{N}\mathcal{Q}_{b}^{P}\mathcal{Q}_{c}^{Q}=\widetilde{\mathcal{Q}}%
_{M\mid \left( abc\right) },  \label{spin=3/2}
\end{equation}
respectively \ sitting in the spin $s=2$ and $s=3/2$ of $SL_{h}\left( 2,%
\mathbb{R}\right) $ (the horizontal indices $a=1,2$ are raised and lowered
with $\epsilon ^{ab}$, with $\epsilon ^{12}\equiv 1$). For clarity's sake,
we report the explicit expressions of the various components of $\mathbf{I}%
_{abcd}$, as well as their relations with the components
of $\widetilde{\mathcal{Q}}_{abc}$ (\ref{spin=3/2}) (the subscripts ``$%
+2,+1,0,-1,-2$'' denote the horizontal helicity of the various components
\cite{FMOSY-1,Irred-1}):
\begin{eqnarray}
\mathbf{I}_{+2} &\equiv &\mathcal{I}_{4}\left( \mathcal{Q}_{1}\right) \equiv
\mathbf{I}_{1111}=\left\langle \widetilde{\mathcal{Q}}_{111},\mathcal{Q}%
_{1}\right\rangle ;  \label{I+2} \\
&&  \notag \\
\mathbf{I}_{+1} &\equiv &\mathbf{I}_{1112}=\left\langle \widetilde{\mathcal{Q%
}}_{111},\mathcal{Q}_{2}\right\rangle =\left\langle \widetilde{\mathcal{Q}}%
_{112},\mathcal{Q}_{1}\right\rangle ;  \label{I+1} \\
&&  \notag \\
\mathbf{I}_{0} &\equiv &\mathbf{I}_{1122}=\left\langle \widetilde{\mathcal{Q}%
}_{112},\mathcal{Q}_{2}\right\rangle =\left\langle \widetilde{\mathcal{Q}}%
_{122},\mathcal{Q}_{1}\right\rangle ;  \label{I0} \\
&&  \notag \\
\mathbf{I}_{-1} &\equiv &\mathbf{I}_{1222}=\left\langle \widetilde{\mathcal{Q%
}}_{122},\mathcal{Q}_{2}\right\rangle =\left\langle \widetilde{\mathcal{Q}}%
_{222},\mathcal{Q}_{1}\right\rangle ;  \label{I-1} \\
&&  \notag \\
\mathbf{I}_{-2} &\equiv &\mathcal{I}_{4}\left( \mathcal{Q}_{2}\right) \equiv
\mathbf{I}_{2222}=\left\langle \widetilde{\mathcal{Q}}_{222},\mathcal{Q}%
_{2}\right\rangle .  \label{I-2}
\end{eqnarray}
Note that in \textit{reducible} symmetric models, due to the \textit{%
semi-simple} nature of $G_{4}$, there are two independent components of $%
\mathbf{I}_{abcd}$ with vanishing horizontal helicity, namely $\mathbf{I}%
^{\prime }$ and $\mathbf{I}^{\prime \prime }$\cite{FMOSY-1}:
\begin{eqnarray}
\mathbf{I}^{\prime } &\equiv &-\frac{1}{2}\text{Tr}_{\eta }\left( \mathbb{T}%
_{11}\mathbb{T}_{22}\right) ; \\
\mathbf{I}^{\prime\prime } &\equiv &-\frac{1}{2}\text{Tr}_{\eta }\left( \mathbb{T}%
_{12}^{2}\right) ,
\end{eqnarray}
and related to
$\mathbf{I}_{0}$ by the relation (4.4) of \cite{FMOSY-1}:
\begin{equation}
\mathbf{I}_{0}=\frac{1}{3}\left( \mathbf{I}^{\prime }+2\mathbf{I}^{\prime
\prime }\right) .  \label{rel-I0-I'-I''}
\end{equation}

Thus, one can consider the following symplectic product of spin $3/2$
horizontal charge tensors:
\begin{equation}
\left\langle \widetilde{\mathcal{Q}}_{abc},\widetilde{\mathcal{Q}}%
_{def}\right\rangle \equiv \widetilde{\mathcal{Q}}_{M\mid abc}\widetilde{%
\mathcal{Q}}_{N\mid def}\mathbb{C}^{MN}.
\end{equation}
\textit{A priori}, $\left\langle \widetilde{\mathcal{Q}}_{abc},\widetilde{%
\mathcal{Q}}_{def}\right\rangle $ should project onto spin $s=3,2,1,0$
irreps. of $SL_{h}\left( 2,\mathbb{R}\right) $ itself; however, due to the
complete symmetry of the $K$-tensor (and to the results of \cite
{Brown-E7,Exc-Reds}), the projections on spin $s=3$ and $1$ do vanish:
\begin{eqnarray}
s &=&3:\left\langle \widetilde{\mathcal{Q}}_{(abc},\widetilde{\mathcal{Q}}%
_{def)}\right\rangle =0;  \label{s=3} \\
s &=&2:\left\langle \widetilde{\mathcal{Q}}_{(ab\mid c},\widetilde{\mathcal{Q%
}}_{d\mid ef)}\right\rangle \epsilon ^{cd}=\frac{2}{3}\mathcal{W}\mathbf{I}%
_{abef};  \label{s=2} \\
s &=&1:\left\langle \widetilde{\mathcal{Q}}_{(a\mid bc},\widetilde{\mathcal{Q%
}}_{de\mid f)}\right\rangle \epsilon ^{bd}\epsilon ^{ce}=0;  \label{s=1} \\
s &=&0:\left\langle \widetilde{\mathcal{Q}}_{abc},\widetilde{\mathcal{Q}}%
_{def}\right\rangle \epsilon ^{ad}\epsilon ^{be}\epsilon ^{cf}=8\mathbf{I}%
_{6},  \label{s=0}
\end{eqnarray}
where the symplectic product $\mathcal{W}$ is defined by (\ref{W}), and in (%
\ref{s=0}) the definition of the sextic horizontal polynomial $\mathbf{I}%
_{6} $ \cite{Irred-1} (given by Eq. (3.24) of \cite{Irred-1}) has been
recalled:
\begin{equation}
\mathbf{I}_{6}\equiv \frac{1}{8}\left\langle \widetilde{\mathcal{Q}}_{abc},%
\widetilde{\mathcal{Q}}_{def}\right\rangle \epsilon ^{ad}\epsilon
^{be}\epsilon ^{cf}=\frac{1}{4}\left\langle \widetilde{\mathcal{Q}}_{111},%
\widetilde{\mathcal{Q}}_{222}\right\rangle +\frac{3}{4}\left\langle
\widetilde{\mathcal{Q}}_{122},\widetilde{\mathcal{Q}}_{112}\right\rangle .
\label{rel-1}
\end{equation}
The complementary relation to (\ref{rel-1}), namely $\frac{1}{4}\left\langle
\widetilde{\mathcal{Q}}_{111},\widetilde{\mathcal{Q}}_{222}\right\rangle -%
\frac{3}{4}\left\langle \widetilde{\mathcal{Q}}_{122},\widetilde{\mathcal{Q}}%
_{112}\right\rangle $ consistently turns out to be proportional (through $%
\mathcal{W}$) to the zero helicity component of $\mathbf{I}_{abcd}$ ;
indeed, by setting $\left( a,b,e,f\right) =\left( 1,1,2,2\right) $ in (\ref
{s=2}), one obtains:
\begin{equation}
\frac{1}{2}\mathbf{I}_{0}\mathcal{W}=\frac{1}{4}\left\langle \widetilde{%
\mathcal{Q}}_{111},\widetilde{\mathcal{Q}}_{222}\right\rangle -\frac{3}{4}%
\left\langle \widetilde{\mathcal{Q}}_{122},\widetilde{\mathcal{Q}}%
_{112}\right\rangle .  \label{s=2,sz=0}
\end{equation}

Furthermore, \textit{at least} in \textit{reducible} symmetric models
(listed in Table 1), the sextic invariant is naturally defined within the $%
\mathbb{T}$-tensor formalism as follows (see Sec. 3 of \cite{FMOSY-1} for
further detail):
\begin{equation}
\mathbf{I}_{6}^{\prime }=-\text{Tr}_{\eta }\left( \mathbb{T}_{11}\mathbb{T}%
_{22}\mathbb{T}_{12}\right) .  \label{I6'}
\end{equation}

As denoted by the prime, $\mathbf{I}_{6}^{\prime }$ (\ref{I6'}) does not
coincide with the $\mathbf{I}_{6}$ given by (\ref{s=0}) (or equivalently by (%
\ref{rel-1})); indeed, \textit{irreducible} and \textit{reducible} symmetric
models, the following relation respectively holds:
\begin{eqnarray}
\mathbf{I}_{6} &=&\mathbf{I}_{6}^{\prime }+\frac{1}{12}\mathcal{W}^{3};
\label{rel-I6-I6'} \\
\mathbf{I}_{6} &=&\mathbf{I}_{6}^{\prime }+\frac{1}{12}\mathcal{W}^{3}+\frac{%
1}{6}\mathcal{X}\mathcal{W}.  \label{rel-I6-I6'-X<>0}
\end{eqnarray}
This can be traced back to a crucial difference (pointed out in Sec. 3 of
\cite{Irred-1}) between \textit{reducible} and \textit{irreducible}
symmetric models, concerning the horizontal invariant polynomial (of degree
four in charges) $\mathcal{X}$. In \textit{reducible} models, $\mathcal{X}$
is defined by Eq. (4.13) of \cite{FMOSY-1}:
\begin{equation}
\mathcal{X}\equiv \text{Tr}_{\eta }\left( \mathbb{T}_{12}^{2}\right) -\text{%
Tr}_{\eta }\left( \mathbb{T}_{11}\mathbb{T}_{22}\right) -\frac{1}{2}\mathcal{%
W}^{2},  \label{Xcall-red}
\end{equation}
and it generally does not vanish. On the other hand, in \textit{irreducible}
models it is defined by Eq. (3.10) of \cite{Irred-1}, and it vanishes
identically:
\begin{equation}
\mathcal{X}_{irred}=0.  \label{Xcall-irred}
\end{equation}
Thus, it is here worth remarking that the relations (\ref{s=3})-(\ref
{s=2,sz=0}) hold both in irreducible models (characterized by (\ref
{Xcall-irred}) and in reducible models (generally having non-vanishing $%
\mathcal{X}$ (\ref{Xcall-red})). On the other hand, in presence of $\mathcal{%
X}\neq 0$, $\mathbf{I}_{6}^{\prime }$ undergoes the renormalization $\mathbf{%
I}_{6}^{\prime }\rightarrow \mathbf{I}_{6}^{\prime }+\frac{1}{6}\mathcal{X}%
\mathcal{W}$, and this explains Eq. (\ref{rel-I6-I6'-X<>0}) from Eq. (\ref
{rel-I6-I6'}).\medskip

Before analyzing and classifying the two-center extremal BH configurations
and the corresponding defining constraints in terms of $G_{4}$- and $\left[
SL_{h}\left( 2,\mathbb{R}\right) \times G_{4}\right] $- invariant
polynomials, in light of previous definitions and findings, we conclude this
Section by pointing out some consequences of the \textit{rank} of a charge
vector, say $\mathcal{Q}_{1}$, on the set of $G_{4}$- and $\left[
SL_{h}\left( 2,\mathbb{R}\right) \times G_{4}\right] $- invariant
polynomials of the two-centered configuration ($\mathcal{Q}_{1},\mathcal{Q}%
_{2}$), both in \textit{irreducible} and \textit{reducible} symmetric
models:
\begin{eqnarray}
\text{\textit{rank}}\left( \mathcal{Q}_{1}\right) &=&3\Rightarrow \mathbf{I}%
_{+2}=0;  \label{rank(Q)=3} \\
&&  \notag \\
\text{\textit{rank}}\left( \mathcal{Q}_{1}\right) &=&2\Rightarrow \widetilde{%
\mathcal{Q}}_{111}=0\Rightarrow \left\{
\begin{array}{l}
\mathbf{I}_{+2}=\mathbf{I}_{+1}=0; \\
\\
\mathbf{I}_{6}^{\prime }=-\frac{1}{2}\mathbf{I}_{0}\mathcal{W}-\frac{1}{12}%
\mathcal{W}^{3}-\frac{1}{6}\mathcal{XW}\Leftrightarrow \mathbf{I}_{6}=-\frac{%
1}{2}\mathbf{I}_{0}\mathcal{W};
\end{array}
\right.  \notag \\
&&  \label{rank(Q)=2} \\
\text{\textit{rank}}\left( \mathcal{Q}_{1}\right) &=&1\Rightarrow \left\{
\begin{array}{l}
\mathbf{I}_{+2}=\mathbf{I}_{+1}=0; \\
\\
\mathbf{I}_{0}=-\frac{1}{6}\mathcal{W}^{2}; \\
\\
\mathbf{I}_{6}^{\prime }=0\Leftrightarrow \mathbf{I}_{6}=-\frac{1}{2}\mathbf{%
I}_{0}\mathcal{W}=\frac{1}{12}\mathcal{W}^{3}; \\
\\
\mathcal{X}=0\text{~(in \textit{reducible} models)}.
\end{array}
\right.  \label{rank(Q)=1}
\end{eqnarray}

\textit{\c{C}a va sans dire} that analogous relations, involving
components of opposite horizontal helicity, hold for
$\mathcal{Q}_{2}$. Eqs. (\ref{rank(Q)=3})-(\ref{rank(Q)=1}) will be
used extensively in Secs. \ref{Red-Analysis} and \ref{Irred-Analysis} (as given by (\ref{Xcall-irred}%
), $\mathcal{X}=0$ identically in \textit{reducible} models). The
non-generic cases of $st^{2}$ and $t^{3}$ $\mathcal{N}=2$, $d=4$ models will
be be considered in Secs. \ref{Red-Analysis-st^2} and \ref
{Irred-Analysis-t^3}, respectively.

\section{\label{Red-Analysis}\textit{Reducible} Models}

As given by Table 1, the \textit{reducible} symmetric $d=4$ supergravity
models\footnote{%
Marginal stability for these models was studied \textit{e.g.} in \cite
{Sen,David-1}.} have the following $d=4$ $U$-duality group:
\begin{equation}
\begin{array}{c}
G_{4}=SL_{v}\left( 2,\mathbb{R}\right) \times SO\left( m,n\right) \\
\Updownarrow \\
SL_{h}\left( 2,\mathbb{R}\right) \times G_{4}\sim SO_{h}^{v}\left(
2,2\right) \times SO\left( m,n\right) ,
\end{array}
~~~m=\left\{
\begin{array}{l}
2~\left( \mathcal{N}=2,~n\in \mathbb{N}\right) ; \\
\\
6~\left( \mathcal{N}=6,~n\in \mathbb{N\cup }\left\{ 0\right\} \right) ,
\end{array}
\right.  \label{G4-red}
\end{equation}
where the isomorphism (see Sec. 8 of \cite{FMOSY-1})
\begin{equation}
SL_{h}\left( 2,\mathbb{R}\right) \times SL_{v}\left( 2,\mathbb{R}\right)
\sim SO_{h}^{v}\left( 2,2\right)  \label{iso-1}
\end{equation}
has been used. The corresponding scalar manifolds thus belong to the
sequence $\mathcal{ST}\left[ m,n\right] $, of particular relevance for
superstring compactifications (see \textit{e.g.} the analysis in Sec. 3.1
and App. C of \cite{N=2-Big}, and Refs. therein).

We now give a complete analysis of all possible two-center charge
configurations $\left( \mathcal{Q}_{1},\mathcal{Q}_{2}\right) $ (with
symplectic product (\ref{W}) $\mathcal{W}\neq 0$, \textit{i.e.} mutually
non-local), by providing for each configuration a ``minimal'' sets of
independent $G_{4}$-invariant and $\left[ SL_{h}\left( 2,\mathbb{R}\right)
\times G_{4}\right] $-invariant polynomials. The analysis will be carried
out in the bare charges $\mathcal{Q}$ basis, by exploiting, for each of the
two centers, the duality-invariant definitions of rank of $\mathcal{Q}$
recalled in Sec. \ref{U-inv-constraints}. The definitions and notation of
\cite{FMOSY-1} are used. The prototype of a generic ($\mathcal{N}=2$)
reducible symmetric model is the $stu$ model, studied in some detail in Sec.
2 of \cite{FMOSY-1}. The non-generic case of the $\mathcal{N}=2$, $d=4$ $%
st^{2}$ model is considered in Sec. \ref{Red-Analysis-st^2}.

\subsection{\label{Generic-Red-Analysis}Generic \textit{Reducible} Models}

\begin{enumerate}
\item  $\left( \mathcal{Q}_{1},\mathcal{Q}_{2}\right) =\left( \text{rank }4,~%
\text{rank }4\right) $. This is the configuration supporting the generic $2$%
-center charge orbits, studied in some detail in \cite{FMOSY-1}. The number
of independent $G_{4}$-invariant is seven, a complete basis can be taken to
be given by Eq. (\ref{p=2-red-duality-basis}). On the other hand, the number
of independent $\left[ SL_{h}\left( 2,\mathbb{R}\right) \times G_{4}\right] $%
-invariant is four, and one can choose a complete basis to be \cite{FMOSY-1}
\begin{equation}
\left[ SL_{h}\left( 2,\mathbb{R}\right) \times G_{4}\right] \text{-inv~}%
\left( \#=4\right) :\text{~}\mathcal{W},~\mathcal{X},~Tr\left( \frak{I}%
^{2}\right) ,~Tr\left( \frak{I}^{3}\right) ,  \label{hor-basis}
\end{equation}

where
\begin{eqnarray}
\text{Tr}\left( \frak{I}^{2}\right)  &=&\mathbf{I}_{+2}\mathbf{I}_{-2}+3%
\mathbf{I}_{0}^{2}-4\mathbf{I}_{+1}\mathbf{I}_{-1}; \\
\text{Tr}\left( \frak{I}^{3}\right)  &=&\mathbf{I}_{0}^{3}+\mathbf{I}_{+2}%
\mathbf{I}_{-1}^{2}+\mathbf{I}_{-2}\mathbf{I}_{+1}^{2}-\mathbf{I}_{+2}%
\mathbf{I}_{-2}\mathbf{I}_{0}-2\mathbf{I}_{+1}\mathbf{I}_{-1}\mathbf{I}_{0}.
\end{eqnarray}

Different choices are of course possible. \textit{E.g.}, equivalent
duality-invariant and horizontal-invariant complete bases respectively read
\begin{eqnarray}
G_{4}\text{-inv} &:&\text{~}\mathcal{W},~\mathcal{X},~\mathbf{I}_{6}^{\prime
},~\mathbf{I}_{\pm 2},~\mathcal{I}_{4}\left( \mathcal{Q}_{1}+\mathcal{Q}%
_{2}\right) ,Tr\left( \frak{I}^{2}\right) ;  \label{alt-1} \\
\left[ SL_{h}\left( 2,\mathbb{R}\right) \times G_{4}\right] \text{-inv} &:&%
\text{~}\mathcal{W},~\mathcal{X},~\mathbf{I}_{6}^{\prime },~Tr\left( \frak{I}%
^{2}\right) .  \label{alt-2}
\end{eqnarray}
The choice (\ref{alt-1})-(\ref{alt-2}) is characterized by the fact that the
horizontal basis is a subset of the duality basis. Furthermore, the duality
basis contains quantities related both to the single-center BH entropy $\pi
\sqrt{\left| \mathcal{I}_{4}\left( \mathcal{Q}_{1}+\mathcal{Q}_{2}\right)
\right| }$ and to the two-centered BH entropy (\ref{S-1+2}). In general, $%
\mathbf{I}_{6}^{\prime }$ is related to the elements of the basis (\ref
{hor-basis}) by means of the polynomial constraint of degree twelve in
charges given by Eq. (5.6) of \cite{FMOSY-1}, which we recall here (see also
the analysis in \cite{Levay-2-center}):
\begin{equation}
\left( \mathbf{I}_{6}^{\prime }\right) ^{2}+\mathcal{WX}\mathbf{I}%
_{6}^{\prime }+Tr\left( \frak{I}^{3}\right) +\frac{1}{12}\mathcal{W}%
^{2}Tr\left( \frak{I}^{2}\right) -\frac{1}{3}\mathcal{X}Tr\left( \frak{I}%
^{2}\right) -\frac{1}{432}\mathcal{W}^{6}+\frac{1}{36}\mathcal{XW}^{4}+\frac{%
5}{36}\mathcal{W}^{2}\mathcal{X}^{2}+\frac{4}{27}\mathcal{X}^{3}=0.
\label{P12=0}
\end{equation}

\item  $\left( \mathcal{Q}_{1},\mathcal{Q}_{2}\right) =\left( \text{rank }3,~%
\text{rank }4\right) $. The complete duality and horizontal bases can
respectively be taken to be
\begin{eqnarray}
G_{4}\text{-inv~}\left( \#=6\right) &:&\text{~}\mathcal{W},~\mathcal{X},~%
\mathbf{I}_{-2},~\mathbf{I}_{\pm 1},~\mathbf{I}_{0}; \\
\left[ SL_{h}\left( 2,\mathbb{R}\right) \times G_{4}\right] \text{-inv~}%
\left( \#=4\right) &:&\text{~}\mathcal{W},~\mathcal{X},~Tr\left( \frak{I}%
^{2}\right) ,~Tr\left( \frak{I}^{3}\right) ,
\end{eqnarray}
where now
\begin{eqnarray}
Tr\left( \frak{I}^{2}\right) &=&3\mathbf{I}_{0}^{2}-4\mathbf{I}_{+1}\mathbf{I%
}_{-1};  \label{r-1} \\
Tr\left( \frak{I}^{3}\right) &=&\mathbf{I}_{0}^{3}+\mathbf{I}_{-2}\mathbf{I}%
_{+1}^{2}-2\mathbf{I}_{+1}\mathbf{I}_{0}\mathbf{I}_{-1}.  \label{r-2}
\end{eqnarray}

\item  $\left( \mathcal{Q}_{1},\mathcal{Q}_{2}\right) =\left( \text{rank }3,~%
\text{rank }3\right) $:
\begin{eqnarray}
G_{4}\text{-inv~}\left( \#=5\right) &:&\text{~}\mathcal{W},~\mathcal{X},~%
\mathbf{I}_{\pm 1},~\mathbf{I}_{0}; \\
\left[ SL_{h}\left( 2,\mathbb{R}\right) \times G_{4}\right] \text{-inv~}%
\left( \#=4\right) &:&\text{~}\mathcal{W},~\mathcal{X},~Tr\left( \frak{I}%
^{2}\right) ,~Tr\left( \frak{I}^{3}\right) ,
\end{eqnarray}
where $Tr\left( \frak{I}^{2}\right) $ is given by (\ref{r-1}), and $Tr\left(
\frak{I}^{3}\right) $ is further simplified to
\begin{equation}
Tr\left( \frak{I}^{3}\right) =\mathbf{I}_{0}\left( \mathbf{I}_{0}^{2}-2%
\mathbf{I}_{+1}\mathbf{I}_{-1}\right) .  \label{r-3}
\end{equation}

\item  $\left( \mathcal{Q}_{1},\mathcal{Q}_{2}\right) =\left( \text{rank }2,~%
\text{rank }4\right) $:
\begin{eqnarray}
G_{4}\text{-inv~}\left( \#=4\right) &:&\text{~}\mathcal{W},~\mathcal{X},~%
\mathbf{I}_{-2},~\mathbf{I}_{-1}; \\
\left[ SL_{h}\left( 2,\mathbb{R}\right) \times G_{4}\right] \text{-inv~}%
\left( \#=2\right) &:&\text{~}\mathcal{W},~\mathcal{X}.
\end{eqnarray}
This case actually splits in two subcases. Indeed, by plugging (\ref
{rank(Q)=2}) into (\ref{P12=0}), this latter factorizes as
\begin{equation}
\left( \mathcal{W}^{2}-4\mathcal{X}+6\mathbf{I}_{0}\right) \left( \mathcal{W}%
^{2}+2\mathcal{X}+6\mathbf{I}_{0}\right) =0,
\end{equation}
thus admitting two solutions:
\begin{eqnarray}
I &:&\left\{
\begin{array}{l}
\mathbf{I}_{0}=-\frac{1}{6}\mathcal{W}^{2}+\frac{2}{3}\mathcal{X}; \\
\\
\mathbf{I}_{6}^{\prime }=-\frac{1}{2}\mathcal{XW}; \\
\\
Tr\left( \frak{I}^{2}\right) =3\mathbf{I}_{0}^{2}=\frac{1}{3}\left( \frac{1}{%
4}\mathcal{W}^{4}+4\mathcal{X}^{2}-2\mathcal{W}^{2}\mathcal{X}\right) ; \\
\\
Tr\left( \frak{I}^{3}\right) =\mathbf{I}_{0}^{3}=\frac{1}{3}\left( -\frac{1}{%
72}\mathcal{W}^{6}+\frac{8}{9}\mathcal{X}^{3}+\frac{1}{2}\mathcal{W}^{4}%
\mathcal{X}-\frac{2}{3}\mathcal{W}^{2}\mathcal{X}^{2}\right) .
\end{array}
\right.  \label{split-1} \\
&&  \notag \\
II &:&\left\{
\begin{array}{l}
\mathbf{I}_{0}=-\frac{1}{6}\mathcal{W}^{2}-\frac{1}{3}\mathcal{X}; \\
\\
\mathbf{I}_{6}^{\prime }=0; \\
\\
Tr\left( \frak{I}^{2}\right) =3\mathbf{I}_{0}^{2}=\frac{1}{3}\left( \frac{1}{%
4}\mathcal{W}^{4}+\mathcal{X}^{2}+\mathcal{W}^{2}\mathcal{X}\right) ; \\
\\
Tr\left( \frak{I}^{3}\right) =\mathbf{I}_{0}^{3}=-\frac{1}{9}\left( \frac{1}{%
24}\mathcal{W}^{6}+\frac{1}{3}\mathcal{X}^{3}+\frac{1}{4}\mathcal{W}^{4}%
\mathcal{X}+\frac{1}{2}\mathcal{W}^{2}\mathcal{X}^{2}\right) .
\end{array}
\right.  \label{split-2}
\end{eqnarray}

\item  $\left( \mathcal{Q}_{1},\mathcal{Q}_{2}\right) =\left( \text{rank }2,~%
\text{rank }3\right) $:
\begin{eqnarray}
G_{4}\text{-inv~}\left( \#=3\right) &:&\text{~}\mathcal{W},~\mathcal{X},~%
\mathbf{I}_{-1}; \\
\left[ SL_{h}\left( 2,\mathbb{R}\right) \times G_{4}\right] \text{-inv~}%
\left( \#=2\right) &:&\text{~}\mathcal{W},~\mathcal{X}.
\end{eqnarray}
The same splitting into subcases $I$ and $II$, as given by (\ref{split-1})
and (\ref{split-2}), occurs.

\item  $\left( \mathcal{Q}_{1},\mathcal{Q}_{2}\right) =\left( \text{rank }2,~%
\text{rank }2\right) $. By recalling (\ref{rank(Q)=2}), the complete duality
and horizontal bases can be taken to coincide:
\begin{equation}
G_{4}\text{-inv~\textit{and~}}\left[ SL_{h}\left( 2,\mathbb{R}\right) \times
G_{4}\right] \text{-inv~}\left( \#=2\right) :\text{~}\mathcal{W},~\mathcal{X}%
.
\end{equation}
The same splitting into subcases $I$ and $II$, as given by (\ref{split-1})
and (\ref{split-2}), occurs. For this configuration, a third subcase $III$
turns out to occur, namely:
\begin{equation}
III:\left\{
\begin{array}{l}
\mathcal{X}=0; \\
\\
\mathbf{I}_{6}^{\prime }=0; \\
\\
\mathbf{I}_{0}=-\frac{1}{6}\mathcal{W}^{2}<0; \\
\\
Tr\left( \frak{I}^{2}\right) =3\mathbf{I}_{0}^{2}=\frac{1}{12}\mathcal{W}%
^{4}; \\
\\
Tr\left( \frak{I}^{3}\right) =\mathbf{I}_{0}^{3}=-\frac{1}{216}\mathcal{W}%
^{6}.
\end{array}
\right.  \label{split-3}
\end{equation}
Thus, for subcase $III$ (\ref{split-3}) $\mathcal{W}$ is the only relevant
polynomial invariant.

\item  $\left( \mathcal{Q}_{1},\mathcal{Q}_{2}\right) =\left( \text{rank }1,~%
\text{rank }4\right) $:
\begin{eqnarray}
G_{4}\text{-inv~}\left( \#=3\right) &:&\text{~}\mathcal{W},~\mathbf{I}_{-2},~%
\mathbf{I}_{-1};  \label{7-1} \\
\left[ SL_{h}\left( 2,\mathbb{R}\right) \times G_{4}\right] \text{-inv~}%
\left( \#=1\right) &:&\text{~}\mathcal{W}.
\end{eqnarray}
Indeed, for this configuration (\ref{rank(Q)=1}) trivially implies (\ref
{split-3}). Thus, the configurations $\left( \mathcal{Q}_{1},\mathcal{Q}%
_{2}\right) =\left( \text{rank }2,~\text{rank }2\right) $ (subcase $III$)
and $\left( \mathcal{Q}_{1},\mathcal{Q}_{2}\right) =\left( \text{rank }1,~%
\text{rank }4\right) $ both implies (\ref{split-3}), but they do differ at
the level of $G_{4}$-invariant polynomials: in the former case only $%
\mathcal{W}$ matters, whereas in the latter case, as given by (\ref{7-1}),
the complete duality basis is three-dimensional.

\item  $\left( \mathcal{Q}_{1},\mathcal{Q}_{2}\right) =\left( \text{rank }1,~%
\text{rank }3\right) $:
\begin{eqnarray}
G_{4}\text{-inv~}\left( \#=2\right) &:&\text{~}\mathcal{W},~\mathbf{I}_{-1};
\\
\left[ SL_{h}\left( 2,\mathbb{R}\right) \times G_{4}\right] \text{-inv~}%
\left( \#=1\right) &:&\text{~}\mathcal{W},
\end{eqnarray}
with Eq. (\ref{split-3}) holding true.

\item  $\left( \mathcal{Q}_{1},\mathcal{Q}_{2}\right) =\left( \text{rank }1,~%
\text{rank }2\right) $. By recalling (\ref{rank(Q)=2}) and (\ref{rank(Q)=1}%
), one obtains that the only relevant polynomial invariant is the symplectic
product $\mathcal{W}$:
\begin{equation}
G_{4}\text{-inv~\textit{and~}}\left[ SL_{h}\left( 2,\mathbb{R}\right) \times
G_{4}\right] \text{-inv~}\left( \#=1\right) :\text{~}\mathcal{W},
\label{only-W}
\end{equation}
with Eq. (\ref{split-3}) holding true.

\item  $\left( \mathcal{Q}_{1},\mathcal{Q}_{2}\right) =\left( \text{rank }1,~%
\text{rank }1\right) $. Eq. (\ref{split-3}) still holds true, and one
obtains that the only relevant polynomial invariant is the symplectic
product $\mathcal{W}$. An example of this configuration is the $D0+D6$
two-constituents solution of the $stu$ model studied in \cite{CS-1} (see
also \cite{Gimon-1}).
\end{enumerate}

\subsection{\label{Red-Analysis-st^2}The $st^{2}$ Model}

We now proceed to consider the non-generic case of the $\mathcal{N}=2$, $d=4$
$st^{2}$ model, which has a rank-$2$ vector multiplets' scalar manifold,
namely $\left[ \frac{SL\left( 2,\mathbb{R}\right) }{U\left( 1\right) }\right]
^{2}$, first ($n=1$) element of the $\mathcal{N}=2$ sequence in Table 1.
Thus, the $U$-duality group is
\begin{equation}
G_{4,st^{2}}=SL_{v}\left( 2,\mathbb{R}\right) \times SO\left( 2,1\right)
\sim SL_{v}\left( 2,\mathbb{R}\right) \times SL\left( 2,\mathbb{R}\right) .
\end{equation}
This model may be obtained as a rank-$2$ truncation (also named \textit{``}$%
st^{2}$\textit{\ degeneration''}; see \textit{e.g.} \cite
{BFGM1,BMOS-1,FMOSY-1}) of the prototype of generic \textit{reducible} $%
\mathcal{N}=2$ symmetric models which, as observed above, is the $stu$ model.

\begin{enumerate}
\item  $\left( \mathcal{Q}_{1},\mathcal{Q}_{2}\right) =\left( \text{rank }4,~%
\text{rank }4\right) $. This is the configuration supporting the generic $2$%
-center orbit, studied (in the BPS case) in some detail in Sec. 6 of
\cite {FMOSY-1}. The number of independent $G_{4}$-invariant is six,
a complete basis can be taken to be \cite{FMOSY-1}
\begin{equation}
\left[ SL_{v}\left( 2,\mathbb{R}\right) \times SO\left( 2,1\right) \right]
\text{-inv~}\left( \#=6\right) :\text{~}\mathcal{W},~\mathcal{X},~\mathbf{I}%
_{\pm 2},~\mathbf{I}_{\pm 1}.
\end{equation}
On the other hand, the number of independent $\left[ SL_{h}\left( 2,\mathbb{R%
}\right) \times G_{4}\right] $-invariant is three, and a complete basis can
be taken to be \cite{FMOSY-1}
\begin{equation}
\left[ SO_{h}^{v}\left( 2,2\right) \times SO\left( 2,1\right) \right] \text{%
-inv~}\left( \#=3\right) :\text{~}\mathcal{W},~\mathcal{X},~Tr\left( \frak{I}%
^{2}\right) .  \label{st^2-hor-inv}
\end{equation}
Two polynomial constraints, of degree sixteen and eight in charges, hold in
the $st^{2}$ model; they are given by Eqs. (6.10)-(6.12) of \cite{FMOSY-1},
which we recall here (see also the analysis in \cite{Levay-2-center}):
\begin{eqnarray}
0 &=&16Tr^{2}\left( \frak{I}^{2}\right) +64Tr\left( \frak{I}^{3}\right)
\mathcal{W}^{2}+\frac{8}{3}Tr\left( \frak{I}^{2}\right) \mathcal{W}^{4}-%
\frac{1}{27}\mathcal{W}^{8}  \notag \\
&&-\frac{32}{3}Tr\left( \frak{I}^{2}\right) \mathcal{X}^{2}+\frac{8}{9}%
\mathcal{W}^{4}\mathcal{X}^{2}+\frac{64}{27}\mathcal{W}^{2}\mathcal{X}^{3}+%
\frac{16}{9}\mathcal{X}^{4};  \label{st^2-constr-1} \\
&&  \notag \\
0 &=&12Tr\left( \frak{I}^{2}\right) -24\mathbf{I}_{6}^{\prime }\mathcal{W}%
-\left( \mathcal{W}^{2}+2\mathcal{X}\right) ^{2}.  \label{st^2-constr-2}
\end{eqnarray}
By means of (\ref{st^2-constr-1}) and (\ref{st^2-constr-2}), $\mathbf{I}%
_{6}^{\prime }$ and $Tr\left( \frak{I}^{3}\right) $ can be expressed in
terms of the horizontal invariants of the basis (\ref{st^2-hor-inv}).

\item  $\left( \mathcal{Q}_{1},\mathcal{Q}_{2}\right) =\left( \text{rank }3,~%
\text{rank }4\right) $:
\begin{eqnarray}
\left[ SL_{v}\left( 2,\mathbb{R}\right) \times SO\left( 2,1\right) \right]
\text{-inv~}\left( \#=5\right) &:&\text{~}\mathcal{W},~\mathcal{X},~\mathbf{I%
}_{-2},~\mathbf{I}_{\pm 1}; \\
\left[ SO_{h}^{v}\left( 2,2\right) \times SO\left( 2,1\right) \right] \text{%
-inv~}\left( \#=3\right) &:&\text{~}\mathcal{W},~\mathcal{X},~Tr\left( \frak{%
I}^{2}\right) ,
\end{eqnarray}
with (\ref{r-1}) and (\ref{r-2}) holding true, as well.

\item  $\left( \mathcal{Q}_{1},\mathcal{Q}_{2}\right) =\left( \text{rank }3,~%
\text{rank }3\right) $:
\begin{eqnarray}
\left[ SL_{v}\left( 2,\mathbb{R}\right) \times SO\left( 2,1\right) \right]
\text{-inv~}\left( \#=4\right) &:&\text{~}\mathcal{W},~\mathcal{X},~\mathbf{I%
}_{\pm 1}; \\
\left[ SO_{h}^{v}\left( 2,2\right) \times SO\left( 2,1\right) \right] \text{%
-inv~}\left( \#=3\right) &:&\text{~}\mathcal{W},~\mathcal{X},~Tr\left( \frak{%
I}^{2}\right) ,
\end{eqnarray}
where $Tr\left( \frak{I}^{2}\right) $ is given by (\ref{r-1}) and $Tr\left(
\frak{I}^{3}\right) $ is further simplified to (\ref{r-3}).

\item  $\left( \mathcal{Q}_{1},\mathcal{Q}_{2}\right) =\left( \text{rank }2,~%
\text{rank }4\right) $:
\begin{eqnarray}
\left[ SL_{v}\left( 2,\mathbb{R}\right) \times SO\left( 2,1\right) \right]
\text{-inv~}\left( \#=4\right) &:&\text{~}\mathcal{W},~\mathcal{X},~\mathbf{I%
}_{-2},~\mathbf{I}_{-1}; \\
\left[ SO_{h}^{v}\left( 2,2\right) \times SO\left( 2,1\right) \right] \text{%
-inv~}\left( \#=2\right) &:&\text{~}\mathcal{W},~\mathcal{X}.
\end{eqnarray}
The splitting into subcases $I$ (\ref{split-1}) and $II$ (\ref{split-2}),
characterizing the generic reducible models (see e.g. point 4 of Sec. \ref
{Red-Analysis}, does \textit{not} occur in the $st^{2}$ model. Indeed, in
such a model the unique solution of the polynomial constraints (\ref
{st^2-constr-1}) and (\ref{st^2-constr-2}) for $\left( \mathcal{Q}_{1},%
\mathcal{Q}_{2}\right) =\left( \text{rank }2,~\text{rank }4\right) $ is
given by Eq. (\ref{split-2}) (namely, only case $II$ holds).

\item  $\left( \mathcal{Q}_{1},\mathcal{Q}_{2}\right) =\left( \text{rank }2,~%
\text{rank }3\right) $:
\begin{eqnarray}
\left[ SL_{v}\left( 2,\mathbb{R}\right) \times SO\left( 2,1\right) \right]
\text{-inv~}\left( \#=3\right) &:&\text{~}\mathcal{W},~\mathcal{X},~\mathbf{I%
}_{-1}; \\
\left[ SO_{h}^{v}\left( 2,2\right) \times SO\left( 2,1\right) \right] \text{%
-inv~}\left( \#=2\right) &:&\text{~}\mathcal{W},~\mathcal{X}.
\end{eqnarray}

\item  $\left( \mathcal{Q}_{1},\mathcal{Q}_{2}\right) =\left( \text{rank }2,~%
\text{rank }2\right) $:
\begin{equation}
\left[ SL_{v}\left( 2,\mathbb{R}\right) \times SO\left( 2,1\right) \right]
\text{-inv~\textit{and~}}\left[ SO_{h}^{v}\left( 2,2\right) \times SO\left(
2,1\right) \right] \text{-inv~}\left( \#=2\right) :\text{~}\mathcal{W},~%
\mathcal{X}.
\end{equation}
For this configuration, there is actually a second subcase given by the
subcase $III$ (\ref{split-3}) of the generic models, in which then $\mathcal{%
W}$ is the only relevant polynomial invariant. In the $st^{2}$ model, (\ref
{split-3}) is realized in the configuration $\left( \mathcal{Q}_{1},\mathcal{%
Q}_{2}\right) =\left( \text{rank }2,~\text{rank }2\right) $ for instance by
setting the charges of the two vector multiplets to coincide; in the ``$%
d=4/d=5$ special coordinates'' symplectic frame, a particular realization of
this is the ``$t^{3}$ degeneration'' (see \textit{e.g.} \cite
{BFGM1,BMOS-1,FMOSY-1}) in which $p^{1}=p^{2}\equiv p$ and $%
q_{1}=q_{2}\equiv q/2$.

\item  $\left( \mathcal{Q}_{1},\mathcal{Q}_{2}\right) =\left( \text{rank }1,~%
\text{rank }4\right) $:
\begin{eqnarray}
\left[ SL_{v}\left( 2,\mathbb{R}\right) \times SO\left( 2,1\right) \right]
\text{-inv~}\left( \#=3\right) &:&\text{~}\mathcal{W},~\mathbf{I}_{-2},~%
\mathbf{I}_{-1}; \\
\left[ SO_{h}^{v}\left( 2,2\right) \times SO\left( 2,1\right) \right] \text{%
-inv~}\left( \#=1\right) &:&\text{~}\mathcal{W}.
\end{eqnarray}
Indeed, also in the $st^{2}$ model Eq. (\ref{split-3}) holds true.

\item  $\left( \mathcal{Q}_{1},\mathcal{Q}_{2}\right) =\left( \text{rank }1,~%
\text{rank }3\right) $. As for the generic reducible models, Eq. (\ref
{split-3}) does hold for $\mathcal{Q}_{1}$, and the complete duality and
horizontal bases can respectively be taken to be
\begin{eqnarray}
\left[ SL_{v}\left( 2,\mathbb{R}\right) \times SO\left( 2,1\right) \right]
\text{-inv~}\left( \#=2\right) &:&\text{~}\mathcal{W},~\mathbf{I}_{-1}; \\
\left[ SO_{h}^{v}\left( 2,2\right) \times SO\left( 2,1\right) \right] \text{%
-inv~}\left( \#=1\right) &:&\text{~}\mathcal{W}.
\end{eqnarray}

\item  $\left( \mathcal{Q}_{1},\mathcal{Q}_{2}\right) =\left( \text{rank }1,~%
\text{rank }2\right) $. Eq. (\ref{split-3}) does hold for $\mathcal{Q}_{1}$,
and the only relevant polynomial invariant is $\mathcal{W}$, as given by Eq.
(\ref{only-W}).

\item  $\left( \mathcal{Q}_{1},\mathcal{Q}_{2}\right) =\left( \text{rank }1,~%
\text{rank }1\right) $. Eq. (\ref{split-3}) does hold for $\mathcal{Q}_{1}$
\textit{and} for $\mathcal{Q}_{2}$, and Eq. (\ref{only-W}) also holds,
\textit{i.e.} the only relevant polynomial invariant is $\mathcal{W}$.
\end{enumerate}

\section{\label{Irred-Analysis}\textit{Irreducible} Models}

We now proceed to consider the \textit{irreducible} symmetric models (see
Table 2), by providing a complete analysis of all possible two-center charge
configurations $\left( \mathcal{Q}_{1},\mathcal{Q}_{2}\right) $ (with
symplectic product (\ref{W}) $\mathcal{W}\neq 0$, \textit{i.e.} mutually
non-local), in the very same way as done for reducible symmetric models in
Sec. \ref{Red-Analysis}. The definitions and notation of \cite{FMOSY-1} and
\cite{Irred-1} are used. The non-generic case of the $\mathcal{N}=2$, $d=4$ $%
t^{3}$ model is considered in Sec. \ref{Irred-Analysis-t^3}.

\subsection{\label{Generic-Irred-Analysis}Generic \textit{Irreducible} Models%
}

\begin{enumerate}
\item  $\left( \mathcal{Q}_{1},\mathcal{Q}_{2}\right) =\left( \text{rank }4,~%
\text{rank }4\right) $. This is the configuration supporting the generic $2$%
-center orbits, studied in some detail in \cite{Irred-1}. As mentioned in
Sec. \ref{2-Center-Relations} and derived in \cite{Irred-1} (see Eq. (3.10)
therein), the quartic horizontal-invariant polynomial $\mathcal{X}_{irred}$
vanishes identically in the class of models under consideration.
Furthermore, (an analogue with $\mathcal{X}_{irred}=0$ of) the polynomial
constraint (\ref{P12=0}) does \textit{not} hold in these models. As
discussed in \cite{Irred-1}, the number of independent $G_{4}$-invariant is
seven, a complete basis can be taken to be given by Eq. (\ref
{p=2-irred-duality-basis}). On the other hand, the number of independent $%
\left[ SL_{h}\left( 2,\mathbb{R}\right) \times G_{4}\right] $-invariant is
four, and one can choose the following complete basis \cite{Irred-1}:
\begin{equation}
\left[ SL_{h}\left( 2,\mathbb{R}\right) \times G_{4}\right] \text{-inv~}%
\left( \#=4\right) :\text{~}\mathcal{W},~\mathbf{I}_{6},~Tr\left( \frak{I}%
^{2}\right) ,~Tr\left( \frak{I}^{3}\right) .
\end{equation}
As for the reducible models analyzed in Sec. \ref{Red-Analysis}, different
choices are of course possible. \textit{E.g.}, an equivalent
duality-invariant complete basis respectively reads
\begin{equation}
G_{4}\text{-inv}:\text{~}\mathcal{W},~\mathbf{I}_{6},~\mathbf{I}_{\pm 2},~%
\mathcal{I}_{4}\left( \mathcal{Q}_{1}+\mathcal{Q}_{2}\right) ,~Tr\left(
\frak{I}^{2}\right) ,~Tr\left( \frak{I}^{3}\right) .  \label{alt-3}
\end{equation}
The choice (\ref{alt-3}) is characterized by the fact that the horizontal
basis is a subset of the duality basis. Furthermore, in this case the
duality basis contains quantities related both to the single-center BH
entropy and to the two-centered BH entropy.

\item  $\left( \mathcal{Q}_{1},\mathcal{Q}_{2}\right) =\left( \text{rank }3,~%
\text{rank }4\right) $:
\begin{eqnarray}
G_{4}\text{-inv~}\left( \#=6\right) &:&\text{~}\mathcal{W},~\mathbf{I}_{6},~%
\mathbf{I}_{-2},~\mathbf{I}_{\pm 1},~\mathbf{I}_{0}; \\
\left[ SL_{h}\left( 2,\mathbb{R}\right) \times G_{4}\right] \text{-inv~}%
\left( \#=4\right) &:&\text{~}\mathcal{W},~\mathbf{I}_{6},~Tr\left( \frak{I}%
^{2}\right) ,~Tr\left( \frak{I}^{3}\right) ,
\end{eqnarray}
where $Tr\left( \frak{I}^{2}\right) $ and $Tr\left( \frak{I}^{3}\right) $
are respectively given by (\ref{r-1}) and (\ref{r-2}).

\item  $\left( \mathcal{Q}_{1},\mathcal{Q}_{2}\right) =\left( \text{rank }3,~%
\text{rank }3\right) $:
\begin{eqnarray}
G_{4}\text{-inv~}\left( \#=5\right) &:&\text{~}\mathcal{W},~\mathbf{I}_{6},~%
\mathbf{I}_{\pm 1},~\mathbf{I}_{0}; \\
\left[ SL_{h}\left( 2,\mathbb{R}\right) \times G_{4}\right] \text{-inv~}%
\left( \#=4\right) &:&\text{~}\mathcal{W},~\mathbf{I}_{6},~Tr\left( \frak{I}%
^{2}\right) ,~Tr\left( \frak{I}^{3}\right) ,
\end{eqnarray}
where $Tr\left( \frak{I}^{2}\right) $ and $Tr\left( \frak{I}^{3}\right) $
are respectively given by (\ref{r-1}) and (\ref{r-3}).

\item  $\left( \mathcal{Q}_{1},\mathcal{Q}_{2}\right) =\left( \text{rank }2,~%
\text{rank }4\right) $:
\begin{eqnarray}
G_{4}\text{-inv~}\left( \#=4\right) &:&\text{~}\mathcal{W},~\mathbf{I}_{-2},~%
\mathbf{I}_{-1},~\mathbf{I}_{0}; \\
\left[ SL_{h}\left( 2,\mathbb{R}\right) \times G_{4}\right] \text{-inv~}%
\left( \#=2\right) &:&\text{~}\mathcal{W},~Tr\left( \frak{I}^{2}\right) =3%
\mathbf{I}_{0}^{2}.
\end{eqnarray}
Furthermore, the limit $\mathbf{I}_{+1}=0$ of (\ref{r-3}) yields
\begin{equation}
Tr\left( \frak{I}^{3}\right) =\mathbf{I}_{0}^{3}.  \label{irred-crit-2}
\end{equation}
Note that in generic irreducible generic models, differently from what
occurs in generic reducible generic models (see \textit{e.g.} point 4 of
Sec. \ref{Generic-Red-Analysis}), the splitting in subcases does \textit{not}
occur.

\item  $\left( \mathcal{Q}_{1},\mathcal{Q}_{2}\right) =\left( \text{rank }2,~%
\text{rank }3\right) $:
\begin{eqnarray}
G_{4}\text{-inv~}\left( \#=3\right) &:&\text{~}\mathcal{W},~\mathbf{I}_{-1},~%
\mathbf{I}_{0}; \\
\left[ SL_{h}\left( 2,\mathbb{R}\right) \times G_{4}\right] \text{-inv~}%
\left( \#=2\right) &:&\text{~}\mathcal{W},~Tr\left( \frak{I}^{2}\right) =3%
\mathbf{I}_{0}^{2},
\end{eqnarray}
with Eq. (\ref{irred-crit-2}) holding true.

\item  $\left( \mathcal{Q}_{1},\mathcal{Q}_{2}\right) =\left( \text{rank }2,~%
\text{rank }2\right) $:
\begin{eqnarray}
G_{4}\text{-inv~}\left( \#=2\right) &:&\text{~}\mathcal{W},~\mathbf{I}_{0};
\\
\left[ SL_{h}\left( 2,\mathbb{R}\right) \times G_{4}\right] \text{-inv~}%
\left( \#=2\right) &:&\text{~}\mathcal{W},~Tr\left( \frak{I}^{2}\right) =3%
\mathbf{I}_{0}^{2},
\end{eqnarray}
with Eq. (\ref{irred-crit-2}) holding true. Equivalently, as done for the
same configuration in generic reducible models as well as in the $st^{2}$
model (see point 6 of Secs. \ref{Generic-Red-Analysis} and \ref
{Red-Analysis-st^2}, respectively), the complete duality and horizontal
bases can be taken to coincide:
\begin{equation}
G_{4}\text{-inv~\textit{and~}}\left[ SL_{h}\left( 2,\mathbb{R}\right) \times
G_{4}\right] \text{-inv~}\left( \#=2\right) :\text{~}\mathcal{W},~Tr\left(
\frak{I}^{2}\right) =3\mathbf{I}_{0}^{2}.
\end{equation}

\item  $\left( \mathcal{Q}_{1},\mathcal{Q}_{2}\right) =\left( \text{rank }1,~%
\text{rank }4\right) $. As for generic reducible models as well as for the $%
st^{2}$ model (see point 7 of Secs. \ref{Generic-Red-Analysis} and \ref
{Red-Analysis-st^2}, respectively), by recalling (\ref{rank(Q)=1}), the
complete duality and horizontal bases can respectively be taken to be
\begin{eqnarray}
G_{4}\text{-inv~}\left( \#=3\right) &:&\text{~}\mathcal{W},~\mathbf{I}_{-2},~%
\mathbf{I}_{-1}; \\
\left[ SL_{h}\left( 2,\mathbb{R}\right) \times G_{4}\right] \text{-inv~}%
\left( \#=1\right) &:&\text{~}\mathcal{W},
\end{eqnarray}
with $\mathbf{I}_{0}=-\frac{1}{6}\mathcal{W}^{2}<0$, and
\begin{eqnarray}
Tr\left( \frak{I}^{2}\right) &=&3\mathbf{I}_{0}^{2}=\frac{1}{12}\mathcal{W}%
^{4};  \label{Cern-1} \\
Tr\left( \frak{I}^{3}\right) &=&\mathbf{I}_{0}^{3}=-\frac{1}{216}\mathcal{W}%
^{6}.  \label{Cern-2}
\end{eqnarray}

\item  $\left( \mathcal{Q}_{1},\mathcal{Q}_{2}\right) =\left( \text{rank }1,~%
\text{rank }3\right) $. Eqs. (\ref{Cern-1})-(\ref{Cern-2}) hold true, and,
as for generic reducible models as well as for the $st^{2}$ model (see point
8 of Secs. \ref{Generic-Red-Analysis} and \ref{Red-Analysis-st^2},
respectively), the complete duality and horizontal bases can respectively be
taken to be
\begin{eqnarray}
G_{4}\text{-inv~}\left( \#=2\right) &:&\text{~}\mathcal{W},~\mathbf{I}_{-1};
\\
\left[ SL_{h}\left( 2,\mathbb{R}\right) \times G_{4}\right] \text{-inv~}%
\left( \#=1\right) &:&\text{~}\mathcal{W}.
\end{eqnarray}

\item  $\left( \mathcal{Q}_{1},\mathcal{Q}_{2}\right) =\left( \text{rank }1,~%
\text{rank }2\right) $. Eqs. (\ref{Cern-1})-(\ref{Cern-2}) hold true, and,
as for generic reducible models as well as for the $st^{2}$ model (see point
9 of Secs. \ref{Generic-Red-Analysis} and \ref{Red-Analysis-st^2},
respectively), the only relevant polynomial invariant is $\mathcal{W}$, as
given by Eq. (\ref{only-W}).

\item  $\left( \mathcal{Q}_{1},\mathcal{Q}_{2}\right) =\left( \text{rank }1,~%
\text{rank }1\right) $. Eqs. (\ref{Cern-1})-(\ref{Cern-2}) hold true, and,
as for generic reducible models as well as for the $st^{2}$ model (see point
10 of Secs. \ref{Generic-Red-Analysis} and \ref{Red-Analysis-st^2},
respectively), the only relevant polynomial invariant is $\mathcal{W}$, as
given by Eq. (\ref{only-W}).
\end{enumerate}

\subsection{\label{Irred-Analysis-t^3}The $t^{3}$ Model}

We now proceed to consider the non-generic case of the $\mathcal{N}=2$, $d=4$
$t^{3}$ model, which has a rank-$1$ vector multiplets' scalar manifold,
namely $\frac{SL\left( 2,\mathbb{R}\right) }{U\left( 1\right) }$ (see the
first line of Table 2. Thus, the $U$-duality group is
\begin{equation}
G_{4}=SL_{v}\left( 2,\mathbb{R}\right) .
\end{equation}
As mentioned above, this model provides a simple yet interesting example,
because it may be obtained both as the rank-$1$ truncation of the \textit{%
reducible} $\mathcal{N}=2$ symmetric models, as well as the first ($q=0$),
non-generic element of the sequence of \textit{irreducible} $\mathcal{N}=2$
symmetric models, which contains the four rank-$3$ magical supergravity
theories \cite{GST}.

\begin{enumerate}
\item  $\left( \mathcal{Q}_{1},\mathcal{Q}_{2}\right) =\left( \text{rank }4,~%
\text{rank }4\right) $. This is the configuration supporting the generic $2$%
-center orbit\footnote{%
This charge configuration has been considered in literature \cite{DM-1,DM-2}
but, within $SL_{v}\left( 2,\mathbb{R}\right) $-invariant polynomials (see
\textit{e.g.} the second possible basis of (\ref{alternatives})), the role
of $\mathbf{I}_{6}$ is not completely clear yet (concerning this, see the
recent study in \cite{Levay-2-center}).}, studied (in the BPS case) in some
detail in Sec. 7 of \cite{FMOSY-1} (see also the comment in the Introduction
of \cite{Irred-1}). The number of independent $G_{4}$-invariant is five, a
complete basis can be taken to be \cite{FMOSY-1}
\begin{equation}
SL_{v}\left( 2,\mathbb{R}\right) \text{-inv~}\left( \#=5\right) :\text{~}%
\mathcal{W},~\mathbf{I}_{6},~\mathbf{I}_{\pm 2},~\mathbf{I}_{+1}.
\end{equation}
However, other equivalent choices read
\begin{equation}
SL_{v}\left( 2,\mathbb{R}\right) \text{-inv~}\left( \#=5\right) :\text{~}%
\left\{
\begin{array}{l}
\mathcal{W},~\mathbf{I}_{6},~\mathbf{I}_{\pm 2},~\mathbf{I}_{0}; \\
or \\
\mathcal{W},~\mathbf{I}_{6},~\mathbf{I}_{\pm 2},~\mathcal{I}_{4}\left(
\mathcal{Q}_{1}+\mathcal{Q}_{2}\right) .
\end{array}
\right.  \label{alternatives}
\end{equation}
On the other hand, the number of independent $\left[ SL_{h}\left( 2,\mathbb{R%
}\right) \times G_{4}\right] $-invariant is two, and a complete basis can be
taken to be \cite{FMOSY-1} (recall isomorphism (\ref{iso-1}))
\begin{equation}
SO_{h}^{v}\left( 2,2\right) \text{-inv~}\left( \#=2\right) :\text{~}\mathcal{%
W},~\mathbf{I}_{6}.  \label{t^3-hor-inv}
\end{equation}
Three polynomial constraints, of degree sixteen, eight and four in charges,
hold in the $t^{3}$ model; they are given by Eqs. (6.12), (7.18), (7.17) and
(7.16) of \cite{FMOSY-1} (see also App. A therein, and the analysis in \cite
{Levay-2-center}), which we recall here:
\begin{eqnarray}
0 &=&16Tr^{2}\left( \frak{I}^{2}\right) +64Tr\left( \frak{I}^{3}\right)
\mathcal{W}^{2}+\frac{8}{3}Tr\left( \frak{I}^{2}\right) \mathcal{W}^{4}-%
\frac{1}{27}\mathcal{W}^{8};  \notag  \label{t^3-1} \\
0 &=&24\mathbf{I}_{6}^{\prime }\mathcal{W}-12Tr\left( \frak{I}^{2}\right) +%
\mathcal{W}^{4};  \notag  \label{t^3-2} \\
\mathcal{X} &=&0.  \label{t^3-3}
\end{eqnarray}
Note that in (\ref{t^3-3}) $\mathcal{X}\equiv \mathcal{X}_{irred}$, and its
identical vanishing consistently characterizes the $t^{3}$ model as a
non-generic irreducible symmetric model. Due to (\ref{t^3-3}), $\mathbf{I}%
_{6}^{\prime }$ and $\mathbf{I}_{6}$ are related through Eq. (\ref
{rel-I6-I6'}). By means of (\ref{t^3-1}), (\ref{t^3-2}) and (\ref{rel-I6-I6'}%
), $Tr\left( \frak{I}^{2}\right) $ and $Tr\left( \frak{I}^{3}\right) $ can
be expressed in terms of the horizontal invariants of the basis (\ref
{t^3-hor-inv}).

\item  $\left( \mathcal{Q}_{1},\mathcal{Q}_{2}\right) =\left( \text{rank }3,~%
\text{rank }4\right) $:
\begin{eqnarray}
SL_{v}\left( 2,\mathbb{R}\right) \text{-inv~}\left( \#=4\right) &:&\text{~}%
\mathcal{W},~\mathbf{I}_{6},~\mathbf{I}_{-2},~\mathbf{I}_{+1}; \\
SO_{h}^{v}\left( 2,2\right) \text{-inv~}\left( \#=2\right) &:&\text{~}%
\mathcal{W},~\mathbf{I}_{6}.
\end{eqnarray}

\item  $\left( \mathcal{Q}_{1},\mathcal{Q}_{2}\right) =\left( \text{rank }3,~%
\text{rank }3\right) $:
\begin{eqnarray}
SL_{v}\left( 2,\mathbb{R}\right) \text{-inv~}\left( \#=3\right) &:&\text{~}%
\mathcal{W},~\mathbf{I}_{6},~\mathbf{I}_{+1}; \\
SO_{h}^{v}\left( 2,2\right) \text{-inv~}\left( \#=2\right) &:&\text{~}%
\mathcal{W},~\mathbf{I}_{6}.
\end{eqnarray}

\item  $\left( \mathcal{Q}_{1},\mathcal{Q}_{2}\right) =\left( \text{rank }2,~%
\text{rank }4\right) $. This charge configuration has been considered in
\cite{BD}. The complete duality and horizontal bases can respectively be
taken to be
\begin{eqnarray}
SL_{v}\left( 2,\mathbb{R}\right) \text{-inv~}\left( \#=2\right) &:&\text{~}%
\mathcal{W},~\mathbf{I}_{-2}; \\
SO_{h}^{v}\left( 2,2\right) \text{-inv~}\left( \#=1\right) &:&\text{~}%
\mathcal{W}.
\end{eqnarray}
By recalling (\ref{t^3-1})-(\ref{t^3-3}), in the $t^{3}$ model the
configuration $\left( \mathcal{Q}_{1},\mathcal{Q}_{2}\right) =\left( \text{%
rank }2,~\text{rank }4\right) $ implies
\begin{equation}
\left\{
\begin{array}{l}
\mathbf{I}_{6}^{\prime }=0\overset{\left( \text{\ref{rel-I6-I6'}}\right) }{%
\Leftrightarrow }\mathbf{I}_{6}=\frac{1}{12}\mathcal{W}^{3}; \\
\\
\mathbf{I}_{0}=-\frac{1}{6}\mathcal{W}^{2}<0; \\
\\
Tr\left( \frak{I}^{2}\right) =\frac{1}{12}\mathcal{W}^{4}; \\
\\
Tr\left( \frak{I}^{3}\right) =-\frac{1}{216}\mathcal{W}^{6}.
\end{array}
\right.  \label{2-4-t^3}
\end{equation}
Consistent with the fact that the $t^{3}$ model can be obtained as a \textit{%
``rank-}$1$\textit{\ degeneration''} (see \textit{e.g.} \cite
{BFGM1,BMOS-1,FMOSY-1}) of the reducible symmetric $\mathcal{N}=2$, $d=4$
models, (\ref{2-4-t^3}) matches the $\mathcal{X}=0$ limit of subcases $I$ (%
\ref{split-1}) and $II$ (\ref{split-2}) or, equivalently, it matches the
subcase $III$ (\ref{split-3}).

\item  $\left( \mathcal{Q}_{1},\mathcal{Q}_{2}\right) =\left( \text{rank }2,~%
\text{rank }3\right) $. The complete duality and horizontal bases can be
taken to coincide, and $\mathcal{W}$ is the only relevant invariant
polynomial in charges:
\begin{equation}
SL_{v}\left( 2,\mathbb{R}\right) \text{-inv~\textit{and~}}SO_{h}^{v}\left(
2,2\right) \text{-inv~}\left( \#=1\right) :\text{~}\mathcal{W}.
\label{W-only-t^3}
\end{equation}

\item  $\left( \mathcal{Q}_{1},\mathcal{Q}_{2}\right) =\left( \text{rank }2,~%
\text{rank }2\right) $. Again, the complete duality and horizontal bases can
be taken to coincide and be given by (\ref{W-only-t^3}).
\end{enumerate}

Since the $t^{3}$ model lacks of an independent doubly-critical ``small''
charge orbit (namely, criticality implies doubly-criticality in this model;
see \textit{e.g.} \cite{FST-last} for a recent account within a $d=3$
timelike-reduced formalism), the cases given by points 7, 8, 9 and 10 are
all missing for this model.

\section{\label{Comment}A Comment on Bound States with Negative Discriminant}

The treatment given in Secs. \ref{Red-Analysis} and
\ref{Irred-Analysis} allows one to discuss in fair generality
two-centered extremal BH compound
states with a given value of $\mathcal{I}_{4}\left( \mathcal{Q}_{1}+\mathcal{%
Q}_{2}\right) $. Here, we would like to comment shortly on some
two-centered charge configurations corresponding to a negative
$\mathcal{I}_{4}\left( \mathcal{Q}_{1}+\mathcal{Q}_{2}\right) $,
\textit{i.e.} to a BH compound that, \textit{regarded as a
single-centered solution}, is a ``large'' non-BPS (extremal) BH.

By recalling the sum rule (\textit{cf.} Eq. (4.7) of \cite{FMOSY-1})
\begin{equation}
\mathcal{I}_{4}\left( \mathcal{Q}_{1}+\mathcal{Q}_{2}\right) =\mathbf{I}%
_{+2}+4\mathbf{I}_{+1}+6\mathbf{I}_{0}+4\mathbf{I}_{-1}+\mathbf{I}_{-2},
\end{equation}
from the analysis of previous Secs. one can single out some
two-charge configurations which necessarily imply
\begin{equation}
\mathcal{I}_{4}\left( \mathcal{Q}_{1}+\mathcal{Q}_{2}\right) =-\mathcal{W}%
^{2}<0.  \label{jj}
\end{equation}
Note that in this case $\mathcal{I}_{4}\left( \mathcal{Q}_{1}+\mathcal{Q}%
_{2}\right) $ becomes horizontal invariant, as well.

For all symmetric \textit{reducible} models (see Table 1), these
configurations are the ones with
\begin{equation}
\text{\textit{reducible}}:~\text{\textit{rank}}\left( \mathcal{Q}_{1},%
\mathcal{Q}_{2}\right) =\left\{
\begin{array}{l}
\left( 2,2\right) _{III}; \\
\left( 1,2\right) ; \\
\left( 1,1\right) ,
\end{array}
\right.
\end{equation}
namely the subcase $III$ of point $6$, and points $9$ and $10$ of
Sec. \ref {Generic-Red-Analysis} (and analogue cases for the
non-generic $st^{2}$ model treated in\ Sec.
\ref{Red-Analysis-st^2}).

For generic \textit{irreducible} models, these configurations are
the ones in which at least one center is \textit{doubly-critical},
namely the ones with
\begin{equation}
\text{\textit{irreducible}}:~\text{\textit{rank}}\left( \mathcal{Q}_{1},%
\mathcal{Q}_{2}\right) =\left\{
\begin{array}{l}
\left( 1,2\right) ; \\
\left( 1,1\right) ,
\end{array}
\right.
\end{equation}
\textit{i.e.} points $9$ and $10$ of Sec.
\ref{Generic-Irred-Analysis}. In the non-generic irreducible $t^{3}$
model, treated in the Sec. \ref {Irred-Analysis-t^3}, the unique
configuration of the kind under
consideration is given by \textit{rank}$\left( \mathcal{Q}_{1},\mathcal{Q}%
_{2}\right) =\left( 2,2\right) $ (point $6$ above), because, as
mentioned in Sec. \ref{Irred-Analysis-t^3}, in such a model
criticality implies doubly-criticality.\medskip

As an illustrative example (which can be realized in heterotic
string theory), let us consider the \textit{``small
}$\mathit{+}$\textit{\ small''} two-centered charge configuration
(in Calabi-Vesentini symplectic frame \cite {CV-sugra}; $\Lambda
=1,....,m+n$)
\begin{equation}
\mathcal{Q}_{1}\equiv \left( p^{\Lambda },0\right)
;~~\mathcal{Q}_{2}\equiv \left( 0,Q_{\Lambda }\right)
\label{Sen-cfg}
\end{equation}
in $d=4$ supergravity coupled to $n$ vector multiplets ($\mathcal{N}=2$ and $%
\mathcal{N}=4$ theories are obtained for $m=2$ and $6$,
respectively),
implying that\footnote{%
In string theory, the quartic invariant $\mathcal{I}_{4}\left( \mathcal{Q}%
\right) $ of \textit{reducible} models is usually named \textit{%
``discriminant''} of the charge vector $\mathcal{Q}$ (see
\textit{e.g.} \cite {Moore}).} \cite{CY,DLR,CT}
\begin{equation}
\mathcal{I}_{4}\left( \mathcal{Q}_{1}+\mathcal{Q}_{2}\right) =\mathcal{I}%
_{4}\left( p,Q\right) =p^{2}Q^{2}-\left( p\cdot Q\right) ^{2}.
\end{equation}
From the analysis of single-centered charge orbits \cite
{CFMZ1,ADFT-FO-1,CFM2}, the charge vector $\mathcal{Q}_{1}$ of (\ref{Sen-cfg}%
) enjoys the following properties, depending on the nature of the
$SO\left( m,n\right) $-vector $p^{\Lambda }$ :
\begin{equation}
\begin{array}{l}
p^{2}\equiv p^{\Lambda }p^{\Sigma }\eta _{\Lambda \Sigma
}>0\Rightarrow
\mathcal{Q}_{1}\text{\textit{rank}}=2\text{, }\frac{1}{2}\text{-BPS}; \\
\\
p^{2}<0\Rightarrow \mathcal{Q}_{1}\text{\textit{rank}}=2\text{,~non-BPS}; \\
\\
p^{2}=0\Rightarrow \mathcal{Q}_{1}\text{\textit{rank}}=1\text{,~}\frac{1}{2}%
\text{-BPS},
\end{array}
\label{case-study}
\end{equation}
and the same holds for $\mathcal{Q}_{2}$ of (\ref{Sen-cfg}), by replacing $%
p^{\Lambda }$ with $Q_{\Lambda }$ ($\eta $ is the $SO\left(
m,n\right) $ metric). By using \textit{e.g.} the formul\ae\ derived
of \cite{FMOSY-1}, one can easily compute that in the heterotic
charge configuration (\ref {Sen-cfg}) there unique two independent
horizontal-invariant polynomials read:
\begin{equation}
\mathcal{W}=-p\cdot Q\neq
0;~~\mathcal{X}=-\frac{1}{2}p^{2}Q^{2}\lesseqgtr 0,
\end{equation}
where the two centers are assumed to have \textit{mutually
non-local} fluxes (and thus it is assumed that $\mathcal{W}\neq 0$).

By considering \textit{both} $\mathcal{Q}_{1}$ and $\mathcal{Q}_{2}$
of configuration (\ref{Sen-cfg}) to be timelike or spacelike (this
corresponds to \textit{rank}$\left(
\mathcal{Q}_{1},\mathcal{Q}_{2}\right) =\left( 2,2\right) _{II}$,
given by subcase $II$ of point $6$ of Sec. \ref
{Generic-Red-Analysis})), and recalling the analysis done in the
second part od Sec. 4 of \cite{Irred-1}, one obtains the case study
reported in Table 3
(in which the cases with $\mathcal{I}_{4}\left( \mathcal{Q}_{1}+\mathcal{Q}%
_{2}\right) =0$ have been disregarded), in which the sign of $\mathcal{I}%
_{4}\left( \mathcal{Q}_{1}+\mathcal{Q}_{2}\right) $ (second column)
is equivalent to the constraint on $\mathcal{X}$ (third column),
because it holds that
\begin{equation}
\mathcal{I}_{4}\left( \mathcal{Q}_{1}+\mathcal{Q}_{2}\right) =\mathcal{I}%
_{4}\left( p,Q\right) =-\mathcal{W}^{2}-2\mathcal{X}.
\label{(2,2)_II}
\end{equation}
\begin{table}[h]
\begin{center}
\begin{tabular}{|c||c|l|c|}
\hline $
\begin{array}{c}
\\
\left( \text{sgn}\left( p^{2}\right) ,\text{sgn}\left( Q^{2}\right)
\right)
\end{array}
$ & $
\begin{array}{c}
\\
\text{sgn}\left(\mathcal{I}_{4}\left(
\mathcal{Q}_{1}+\mathcal{Q}_{2}\right) \right)
\end{array}
~$ & $
\begin{array}{l}
\\
\text{constraint}
\end{array}
$ & $
\begin{array}{c}
\\
\text{horizontal} \\
\text{orbit~}\mathcal{O}/SO_{h}^{v}\left( 2,2\right)
\end{array}
$ \\ \hline\hline $
\begin{array}{c}
\\
\left( +,+\right) \\
~
\end{array}
$ & $
\begin{array}{c}
\\
+ \\
~
\end{array}
$ & \multicolumn{1}{|c|}{$
\begin{array}{c}
\\
\mathcal{X}<-\frac{1}{2}\mathcal{W}^{2} \\
~
\end{array}
$} & $\frac{SO\left( m,n\right) }{SO(m-2,n)}$ \\ \hline $
\begin{array}{c}
\\
\left( +,+\right) \\
~
\end{array}
$ & $
\begin{array}{c}
\\
- \\
~
\end{array}
$ & \multicolumn{1}{|c|}{$
\begin{array}{c}
\\
\mathcal{X}>-\frac{1}{2}\mathcal{W}^{2}
\end{array}
$} & $\frac{SO\left( m,n\right) }{SO(m-1,n-1)}$ \\ \hline $
\begin{array}{c}
\\
\left( -,-\right) \\
~
\end{array}
$ & $
\begin{array}{c}
\\
+ \\
~
\end{array}
$ & \multicolumn{1}{|c|}{$
\begin{array}{c}
\\
\mathcal{X}<-\frac{1}{2}\mathcal{W}^{2} \\
~
\end{array}
$} & $\frac{SO\left( m,n\right) }{SO(m,n-2)}$ \\ \hline $
\begin{array}{c}
\\
\left( -,-\right) \\
~
\end{array}
$ & $
\begin{array}{c}
\\
- \\
~
\end{array}
$ & \multicolumn{1}{|c|}{$
\begin{array}{c}
\\
\mathcal{X}>-\frac{1}{2}\mathcal{W}^{2}
\end{array}
$} & $\frac{SO\left( m,n\right) }{SO(m-1,n-1)}$ \\ \hline $
\begin{array}{c}
\\
\left( +,-\right) \\
~
\end{array}
$ & $
\begin{array}{c}
\\
- \\
~
\end{array}
$ & \multicolumn{1}{|c|}{$
\begin{array}{c}
\\
\mathcal{X}>0
\end{array}
$} & $\frac{SO\left( m,n\right) }{SO(m-1,n-1)}$ \\ \hline $
\begin{array}{c}
\\
\left( -,+\right) \\
~
\end{array}
$ & $
\begin{array}{c}
\\
- \\
~
\end{array}
$ & \multicolumn{1}{|c|}{$
\begin{array}{c}
\\
\mathcal{X}>0
\end{array}
$} & $\frac{SO\left( m,n\right) }{SO(m-1,n-1)}$ \\ \hline
\end{tabular}
\end{center}
\caption{Two-centered charge configuration of the type
(\ref{Sen-cfg}) with \textit{rank}$\left(
\mathcal{Q}_{1},\mathcal{Q}_{2}\right) =\left( 2,2\right) _{II}$ in
\textit{reducible} symmetric models ($\mathcal{N}=2$ and
$\mathcal{N}=4$ supergravity theories are obtained for $m=2$ and
$m=6$,
respectively). The cases with $\mathcal{I}_{4}\left( \mathcal{Q}_{1}+%
\mathcal{Q}_{2}\right) =0$ are not considered. }
\end{table}
On the other hand, when $p^{\Lambda }$ \textit{and/or} $Q_{\Lambda }$ of (%
\ref{Sen-cfg}) are lightlike (or, equivalently, when
$\mathcal{Q}_{1}$ and/or $\mathcal{Q}_{2}$ are \textit{rank} $1$;
see Eq. (\ref{case-study})), namely in the \textit{mutually
non-local} cases
\begin{equation}
\begin{array}{ccc}
~p^{2}>0, & Q^{2}=0: & \text{\textit{rank}}=\left( 2,1\right) ; \\
p^{2}=0, & Q^{2}>0: & \text{\textit{rank}}=\left( 1,2\right) ; \\
~p^{2}=0, & Q^{2}=0: & \text{\textit{rank}}=\left( 1,1\right) ,
\end{array}
\label{case-study-2}
\end{equation}
$\mathcal{I}_{4}\left( \mathcal{Q}_{1}+\mathcal{Q}_{2}\right) $ is
strictly negative (because $\mathcal{X}=0$), and it is given by Eq.
(\ref{jj}). Note
that for the cases (\ref{case-study-2}) $\mathcal{I}_{4}\left( \mathcal{Q}%
_{1}+\mathcal{Q}_{2}\right) =0$ is equivalent to \textit{mutually
local} centers.

Thus, the two-centered charge configurations $\left( \mathcal{Q}_{1},%
\mathcal{Q}_{2}\right) $ of the heterotic type (\ref{Sen-cfg}) given
by the second, fourth, fifth and sixth line of Table 3, and by Eq.
(\ref {case-study-2}), are all characterized by the corresponding BH
compound that, \textit{regarded as a single-centered (extremal) BH},
is ``large'' and non-BPS, with a negative quartic duality invariant
$\mathcal{I}_{4}\left( \mathcal{Q}_{1}+\mathcal{Q}_{2}\right) $.
These configurations are (the semiclassical limit, with real,
continuous charges of) some of the cases
recently analyzed by Sen in\footnote{%
In the notation of \cite{Sen-2-center}, $m\equiv R$ and $n\equiv
L$.} \cite {Sen-2-center} in $\mathcal{N}=4$, $d=4$ supergravity;
furthermore, the case \textit{rank}$\left(
\mathcal{Q}_{1},\mathcal{Q}_{2}\right) =\left(
1,1\right) $ encompasses the $D0+D6$ configuration in the $\mathcal{N}=2$ $%
stu$ model, which has been studied in \cite{CS-1}. Since both
(\ref{jj}) and (\ref{(2,2)_II}) are manifestly horizontal-invariant
expression, the most general supporting two-centered configurations
can be computed by acting on the relevant case of the heterotic
configuration (\ref{Sen-cfg}) with a generic $\left[
SO_{h}^{v}\left( 2,2\right) \times SO\left( m,n\right) \right]
$-transformation (also, recall (\ref{iso-1})).

\section*{Acknowledgments}

A. M. would like to thank Leron Borsten for interesting discussions, the
Department of Theoretical Physics, University of Torino, for kind
hospitality, and the ERC Advanced Grant no. 226455, \textit{``Supersymmetry,
Quantum Gravity and Gauge Fields''} (\textit{SUPERFIELDS}) for financial
support.

A. Y. would like to thank CERN Theory Division for kind hospitality.

The work of A. C., S. F. and A. Y. is supported by the ERC Advanced Grant no. 226455 \textit{%
SUPERFIELDS}. S. F. is supported in part by DOE Grant
DE-FG03-91ER40662.

\end{document}